\documentclass[11pt]{iopart}
\usepackage{comment}
\usepackage{enumerate}
\usepackage{amsmath,amssymb}
\usepackage{dsfont}
\usepackage{braket}
\usepackage{graphicx}
\usepackage[usenames,dvipsnames]{color}
\usepackage{tensor}
\usepackage{empheq}
\usepackage{capt-of}
\usepackage[normalem]{ulem} 
\usepackage{soul} %
\usepackage{MnSymbol}

\usepackage[colorlinks,bookmarks=false,citecolor=NavyBlue,linkcolor=OliveGreen,urlcolor=blue]{hyperref}
\newcommand{\be}{\begin{equation}}
\newcommand{\ee}{\end{equation}}
\newcommand{\ba}{\begin{aligned}}
\newcommand{\ea}{\end{aligned}}
\newcommand{\bw}{\begin{widetext}}
\newcommand{\ew}{\end{widetext}}

\newcommand{\bea}{\begin{eqnarray}}
\newcommand{\eea}{\end{eqnarray}}

\def\doi{http://dx.doi.org/}

\begin{document}
\title{Bound-state confinement after trap-expansion dynamics in  integrable systems}
\author{Leonardo Biagetti}
\address{Laboratoire de Physique Th\'eorique et Mod\'elisation, CNRS UMR 8089,
CY Cergy Paris Universit\'e, 95302 Cergy-Pontoise Cedex, France}
\author{Vincenzo Alba}
\address{Dipartimento di Fisica dell' Universit\`a di Pisa and INFN, Sezione di Pisa, I-56127
	Pisa, Italy}

\begin{abstract}
	Integrable systems possess stable families of quasiparticles, which are composite objects (bound states) 
	of elementary excitations. Motivated by recent quantum computer experiments, we investigate 
	bound-state transport in the spin-$1/2$ anisotropic Heisenberg chain ($XXZ$ chain). Specifically, 
	we consider the sudden vacuum expansion of a finite region $A$ prepared in a non-equilibrium state. 
	In the hydrodynamic regime, if interactions are strong enough, bound states remain confined 
	in the initial region. 
	Bound-state confinement persists until the density of unbound excitations remains finite in the bulk 
	of $A$. Since region $A$ is finite, at asymptotically long times bound states are 
	``liberated'' after the ``evaporation'' of  the unbound excitations. 
	Fingerprints of confinement are visible in the space-time profiles of local 
	spin-projection operators. 
	To be specific, here we focus on the expansion of the $p$-N\'eel states, which are obtained by repetition 
	of a unit cell with $p$ up spins followed by $p$ down spins. Upon increasing $p$, the bound-state content 
	is enhanced. In the limit $p\to\infty$ one obtains the domain-wall initial state. 
	We show that for $p<4$, only bound states with $n>p$ are confined at large chain anisotropy. 
	For $p\gtrsim 4$, also bound states with $n=p$  are confined, consistent with the absence of transport in 
	the limit $p\to\infty$. 
	The scenario of bound-state confinement leads to a hierarchy of timescales at which bound states 
	of different sizes are liberated, which is also reflected in the dynamics of the von Neumann entropy. 
\end{abstract}

\maketitle
\noindent 

\section{Introduction}
\label{sec:intro}

Integrable quantum many-body systems provide an ideal playground to discover exotic behaviors both at 
equilibrium and out-of-equilibrium. As recognized already by Bethe~\cite{bethe1931zur}, one of the 
hallmarks of interacting many-body systems is the formation of stable multiparticle bound states. 
Bound states play a pivotal role in the 
equilibrium and out-of-equilibrium integrable systems, as they enter prominently in the 
Thermodynamic Bethe Ansatz~\cite{takahashi1999thermodynamics} approach. Since the birth of 
Bethe Ansatz, bound states attracted growing attention, also 
experimentally~\cite{haller2009realization,fukuhara2013microscopic,wang2018experimental,bera2020dispersion,scheie2022quantum,kranzl2023observation}. Furthermore, bound states are key ingredients  behind exotic transport behaviors, 
such as superdiffusion~\cite{ilievski2021superuniversality}. Very recently, 
they have been observed in quantum-computing platforms~\cite{morvan2022formation,surace2024robustness}. 
We should mention that while bound states should be present also in generic, i.e., non integrable 
models, integrability ensures that they are stable excitations. 

Here we investigate the dynamics of bound states after quenches from inhomogeneous initial states in 
the Heisenberg $XXZ$ spin chain, which is a prototypical integrable system~\cite{takahashi1999thermodynamics}. 
An important property of integrable systems is that they possess stable quasiparticles, which are  
labelled by  a set of complex numbers called rapidities~\cite{takahashi1999thermodynamics}. 
For instance, the quasiparticles of the Heisenberg spin chain  
resemble standard magnon excitations and bound states of magnons. 
In contrast with standard magnons, however, the excitations of  Bethe Ansatz systems are 
interacting. Crucially, quasiparticles and their rapidities are preserved during the dynamics because 
integrability allows only elastic scattering. 
This property allows to understand the dynamics of local observables, and of entanglement~\cite{alba2017entanglement}, 
in terms of the ballistic propagation of the quasiparticles. This observation is 
at the heart of the so-called Generalized Hydrodynamics ($GHD$) approach~\cite{bertini2016transport,castro-alvaredo2016emergent}. 
 
\begin{figure*}[t]
\includegraphics[width=\linewidth]{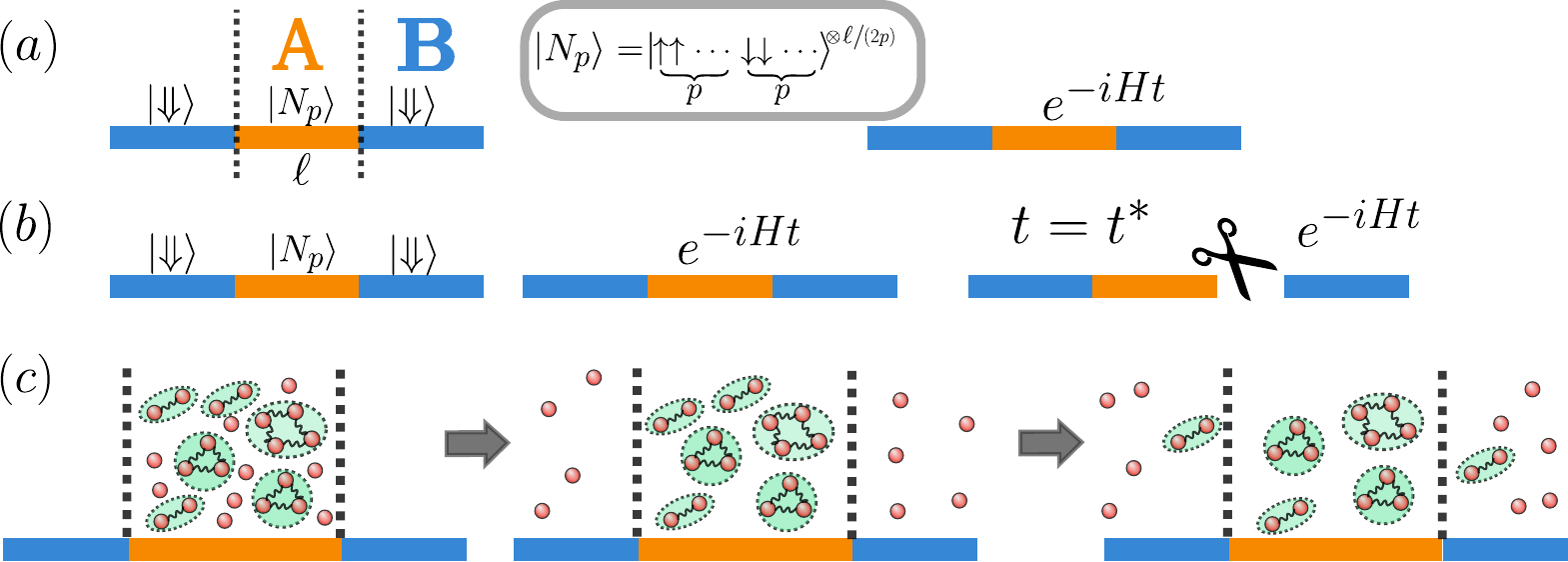}
\caption{ Quench protocols considered in this work. Initial states.  (a) Expansion of $p$-N\'eel states in the 
 vacuum. 
 At $t=0$ a chain prepared in the $p$-N\'eel state is joined with two chains prepared in the 
 state $|\Downarrow\rangle$ with all the spins down. 
 The $p$-N\'eel states $|N_p\rangle$ are obtained by repetition of 
 a unit cell with $p$ up spins followed by $p$ down spins.
 At $t>0$ the full system evolves under the 
 $XXZ$ chain Hamiltonian. 
 (b) The ``cut'' quench. The initial state is the same as in (a). For 
 $0<t\le t^*$ the full chain is evolved under the $XXZ$ Hamiltonian. At $t>t^*$ only part $B$ 
 of the chain is evolved. Here $t^*$ is chosen large enough for the $GHD$ description to be  
 valid, but not too large, to ensure that the initial $p$-N\'eel state is not fully ``melted''. 
 (c) Bound-state confinement. The initial state of region $A$ gives rise to a thermodynamic macrostate 
 containing bound states of different size $n$. The bound states with 
 $n\ge 2$ remain confined in region $A$ until the magnons, i.e, quasiparticles with 
 $n=1$,  start ``evaporating'' at $t\sim\ell$. After the magnons evaporated, 
 the two-particle bound states start leaving region $A$. 
}
\label{fig0:quench}
\end{figure*}

Our out-of-equilibrium protocol is depicted in Fig.~\ref{fig0:quench}:  A 
region $A$ of size $\ell$ is embedded in an infinite chain, and  
is prepared in a product state. The rest of the chain is in  the 
ferromagnetic state with all the spins down. The ferromagnetic state is the 
vacuum state for the Bethe Ansatz quasiparticles. The full chain then evolves under the 
$XXZ$ Hamiltonian. Since quasiparticles are initially 
present only in region $A$, whereas the remaining part of the chain is in the vacuum state, our 
protocol is akin to the trap-expansion protocol routinely employed in cold atom experiments. For this 
reason, in the following we often refer to the protocol in Fig.~\ref{fig0:quench} as trap-expansion or expansion protocol. 
The initial state is not globally an eigenstate of the $XXZ$ model, implying that 
the system evolves with time. 
Specifically, the bulk of region $A$ undergoes a nontrivial evolution from time $t=0^+$, whereas the 
bulk of the remaining part of the chain initially does not evolve, because it is in the vacuum state.

The dynamics of region $A$ is understood in terms of the quasiparticle 
excitations~\cite{takahashi1999thermodynamics} of the model.  
For times $t\ll \ell$, provided that $t$ is long enough for the system to reach local equilibrium, 
local observables in 
the bulk of $A$ are described by a thermodynamic \emph{macrostate}, which is a 
Generalized Gibbs Ensemble~\cite{calabrese2016introduction} ($GGE$). 
Moreover, the initial inhomogeneity that is present 
at the interface between region $A$ 
and the rest (see Fig.~\ref{fig0:quench}) ``melts'' during the dynamics. 
This is due to the fact that the quasiparticles that initially are present only in $A$ 
spread in the rest of the chain. At long times, different regions around the edges of $A$ reach 
different equilibrium macrostates, giving rise to a space-time dependent 
$GGE$.  This inhomogeneous $GGE$ can be  obtained within the $GHD$ 
approach~\cite{bertini2016transport,castro-alvaredo2016emergent}. 
The main idea of $GHD$ (see section~\ref{sec:ghd}) is that the space-time dependent densities of quasiparticles 
satisfy a continuity equation, reflecting that quasiparticles do not decay and preserve their rapidity. 
Moreover, since the scattering between different quasiparticles 
is elastic, its net effect  is to dress the ``bare'' velocities of the 
quasiparticles. In turn, the dressed velocities, which appear in the continuity equation, 
depend on the space-time-dependent densities. This means that the continuity equation 
has to be solved self-consistently. 

Crucially, the effect of the scatterings can be so dramatic that the sign of the 
velocities  changes. Typically, this is more striking for the bound states,  
because their bare velocities (see section~\ref{sec:conf}) are 
generically smaller than those of the magnons. 
As we will show, this leads to bound-state confinement, meaning that the 
transport of bound states from region $A$ to  region $B$ is completely suppressed at ballistic level,  i.e. at long times $t$ and large distances $x$ from 
the interface between $A$ and the rest, with the ratio $\zeta:=x/t$ fixed. In the rest of the paper, 
the confinement of bound states must always be intended in this limit.

Intuitively, as the bound states approach the interface between $A$ and the rest, they scatter 
with the magnons, and, as a consequence, they are ``pushed back'' in $A$. This 
leads to an accumulation of bound states in $A$. From now on we refer to this behavior, i.e., to the fact that 
the bound states remain trapped in region $A$ at ballistic level, as bound state 
confinement. This was observed already in Ref.~\cite{alba2018entanglementand} 
(see also~\cite{alba2019towards}). Moreover, the sign change of the bound-state velocities  
was analyzed at the level of the space-time trajectories 
in Ref.~\cite{alba2019entanglement}, and it has striking effects on the 
entanglement dynamics~\cite{alba2019entanglement,alba2021generalized}. 
Besides confinement, the presence of  bound states  gives rise to several intriguing 
transport behaviors, such as nonanaliticities in the 
profile of local observables~\cite{piroli2017transport}. 
%
%
Finally, we should mention that bound-state confinement gives 
rise to the so-called \emph{quantum distillation}~\cite{heidrich-meisner2009quantum,xia2015quantum,scherg2018nonequilibrium}. 
It has been observed in Ref.~\cite{heidrich-meisner2009quantum} that after the sudden expansion of interacting fermions if the initial state 
has a strong admixture of double occupancies, the doublons group together forming a band-insulating state trapped  at the center of the system. 
This state has low-entropy. This protocol has been investigated in cold-atom experiments in~\cite{xia2015quantum,scherg2018nonequilibrium}.

In this work we provide a thorough investigation of bound-state confinement in the $XXZ$ chain at 
anisotropy $\Delta>1$ by combining the time-dependent Density Matrix Renormalization Group~\cite{paeckel2019time} ($tDMRG$) 
with state-of-the-art integrability methods~\cite{ilievski2016string,ilievski2016quasilocal}. 
For $\Delta>1$ the $XXZ$ chain possesses bound states of an arbitrary number of magnons.  
We show that all the multiparticle bound states remain confined in $A$, provided that interactions, i.e., the 
anisotropy $\Delta$, are strong enough, and there is a finite density of 
magnonic quasiparticles, i.e., unbound ones, in region $A$. Confinement is present at large 
$\Delta$, and it can be understood perturbatively in the limit $\Delta\to\infty$. Upon lowering 
$\Delta$ there is a threshold  below which the lighter bound states, i.e., formed by two 
magnons, are not confined.  This means that  two-particle bound states  
are allowed to leave region $A$ (see Fig.~\ref{fig0:quench}). This is in agreement with the results of 
Ref.~\cite{alba2018entanglementand} and~\cite{alba2019towards} for the specific cases of N\'eel and Majumdar-Ghosh 
initial states. Moreover, for typical initial states of region $A$ there is a hierarchical emission of 
bound states, meaning that heavier bound states are ``liberated'' at lower $\Delta$. 
While confinement can be understood perturbatively, the deconfinement happens near $\Delta=1$, i.e., away from 
the perturbative regime. This means that the deconfinement pattern of the different bound states 
depends on the details of the initial state, and it cannot be addressed perturbatively. 
Moreover, it is important to stress that if the size $\ell$ of region $A$ is finite, confinement 
can be observed only at intermediate times $t\ll \ell$. 
As it will explained in Section \ref{sec:conf}, intermediate time $t\ll \ell$ are both a requirement for the applicability of the bipartite protocol in GHD and to ensure that the central region has a finite density of excitations.
Indeed, at times $t\sim \ell$ the density of 
magnons in the bulk of $A$ starts to be depleted. This implies that the effect of the scattering 
between magnons and bound states, and hence the dressing of bound-state velocity, weakens. 
At long-enough times, after the magnons ``evaporated'' from $A$, deconfinement of the bound states happens 
(see Fig.~\ref{fig0:quench} (c)). 
Following Ref.~\cite{ganahl2012observation}, we employ the space-time profiles of the 
spin projectors $P_{x,\uparrow^{\otimes n}}$ to witness bound-state confinement. 
$P_{x,\uparrow^{\otimes n}}$ are sensitive to the presence of a string of $n$ up 
spins on consecutive sites starting from $x$. 
We also confirm the confinement scenario by considering a ``cut'' quench (see Fig.~\ref{fig0:quench} (b)) 
in which the expansion is stopped after a time $t^*$, and for $t>t^*$ only part $B$ of the chain evolves. 
The stopping time is chosen such that the density of magnons in $A$ is finite, i.e., before the bound-state 
deconfinement starts. 
If interactions  
are strong enough to permit bound-state confinement, the expectation values of the 
projectors $P_{\uparrow^{\otimes n}}$ do not show any signal at any $t>t^*$ in region $B$, since there are no 
bound states outside of $A$. 

In this work, we focus on the expansion of  the $p$-N\'eel states, which are 
``fattened'' version of the N\'eel state. They are defined by repetition of a unit cell containing 
$p$ up spins followed by $p$ down spins (see Fig.~\ref{fig0:quench} (a)). An interesting feature of the $p$-N\'eel state 
is that by changing $p$, one can change the bound-state content. Precisely, states with larger $p$  contain 
larger bound states.  
We observe the following pattern  for bound state confinement. If region $A$ 
(see Fig.~\ref{fig0:quench}) is prepared in the N\'eel state (with $p=1$), 
at large enough $\Delta\gtrsim 1.4$, all the bound states are confined, as observed in Ref.~\cite{alba2019towards}. 
Interestingly, for $2\le p < 4$, 
only bound states with size $n>p$ are confined, whereas for  $p\ge 4$ bound states of size $n=p$  
are also confined. 

Finally, we show that bound-state confinement and deconfinement are sharply reflected in the dynamics of 
the entanglement entropy. First, within the quasiparticle picture for 
entanglement dynamics~\cite{calabrese2005evolution,fagotti2008evolution,alba2017entanglement}, 
entanglement is created in region $A$. Transport between $A$ and 
the rest does not generate new entanglement, but entanglement is simply ``transported'' from $A$ to the 
rest~\cite{alba2021generalized}. We focus on the full expansion 
of the N\'eel state. We numerically show that the hierarchical emission of bound states of different sizes is 
reflected in a multistage growth of the entanglement entropy.

The manuscript is organized as follows. In section~\ref{sec:prot} we introduce the $XXZ$ spin chain and 
the out-of-equilibrium protocols. In section~\ref{sec:tba} we review the Thermodynamic Bethe Ansatz ($TBA$) of the 
model. In section~\ref{sec:ghd} we discuss the Generalized Hydrodynamics ($GHD$) approach. In section~\ref{sec:conf} 
by using the $GHD$ approach, we investigate bound-state confinement. In particular, in section~\ref{sec:ghd-num} 
we provide numerical $GHD$ results confirming bound-state confinement after the expansion of the $p$-N\'eel states. 
In section~\ref{sec:numerics}, we discuss numerical $tDMRG$ results for the expansion dynamics. In particular, 
in section~\ref{sec:check-ghd} we discuss the validity of the $GHD$ approach. 
In section~\ref{sec:full} and section~\ref{sec:cut-quench} we provide $tDMRG$ data for the full-expansion dynamics 
and the cut protocol (see Fig.~\ref{fig0:quench} (a) and (b)), respectively. In section~\ref{sec:ent}, we focus on 
the entanglement entropy after the full-expansion protocol. We conclude and discuss future 
directions in section~\ref{sec:concl}. In~\ref{app:tba_gge} we discuss the $TBA$ derivation of the 
$GGE$ for homogeneous quenches. In~\ref{sec:bond} we discuss the convergence of the $tDMRG$ method.

\section{Model and quantum transport protocols}
\label{sec:prot}

Here we focus on the spin-$1/2$ anisotropic Heisenberg chain ($XXZ$ chain) defined 
by the Hamiltonian 
\begin{equation}
	\label{eq:ham}
	H=\sum_{j=1}^{L}\sigma_j^x\sigma_{j+1}^x+\sigma_{j}^y\sigma_{j+1}^y
	+\Delta(\sigma_j^z\sigma_{j+1}^z-1), 
\end{equation}
where $\sigma_j^{\alpha}$ are Pauli matrices, $\Delta$ is the co-called anisotropy parameter, and 
$L$ is the length of the chain. We focus on the situation with $\Delta\ge 1$.  We 
employ periodic boundary conditions. However, since our results do 
not depend on boundary conditions we employ open boundary conditions in the 
$tDMRG$ numerical simulations~\cite{paeckel2019time}. For both periodic and open 
boundary condition, the $XXZ$ is integrable by the Bethe Ansatz~\cite{takahashi1999thermodynamics}. 

Here we consider the expansion of ``fat'' N\'eel states (see Fig.~\ref{fig0:quench} (a)). They 
are obtained by repetition of a unit cell containing $2p$ spins, with $p$ spins being up and 
$p$ down. We denote these $p$-N\'eel states as $|N_p\rangle$. The generic $p$-N\'eel state reads as 
\begin{equation}
	\label{eq:pneel}
	|N_p\rangle:=|\uparrow^{\otimes p} \downarrow^{\otimes p}\rangle^{\otimes \ell/(2p)}, 
\end{equation}
where $\otimes$ denotes the tensor product.

In the limit $p\to \infty$ the 
$p$-N\'eel states converge to the domain-wall state. Interestingly, for $\Delta>1$ transport is suppressed 
after the quench from the domain-wall state~\cite{gobert-2005}. At $\Delta=1$ transport is neither ballistic nor diffusive, 
but a superdiffusive dynamics is expected~\cite{ilievski2018superdiffusion}.

The quench protocols that we employ  are described in Fig.~\ref{fig0:quench} $(a)$ and $(b)$. 
In both $(a)$ and $(b)$ the initial state at $t=0$ is obtained by embedding a 
chain $A$ of size $\ell$ prepared in the state $|N_p\rangle$ in the vacuum state, 
i.e., the ferromagnetic state $|\Downarrow\rangle$ with all the spins down 
$|\Downarrow\rangle=|\downarrow\downarrow\cdots\rangle$. 
Importantly, since $|\Downarrow\rangle$ is an eigenstate of~\eqref{eq:ham} for any $\Delta$, only 
region $A$ (see Fig.~\ref{fig0:quench}) exhibits a nontrivial dynamics at initial time $t=\epsilon\to0$. 

In protocol $(a)$ the full system undergoes unitary dynamics under the Hamiltonian~\eqref{eq:ham}. 
In the second protocol (Fig.~\ref{fig0:quench} $(b)$) we evolve the full system up to a time $t^*$. 
As the dynamics progresses, the initial $p$-N\'eel state ``melts''.  
For $t>t^*$ the evolution happens under the Hamiltonian~\eqref{eq:ham} restricted to the region $B$ 
on the right of $|N_p\rangle$ (see Fig.~\ref{fig0:quench}), 
whereas part $A$ of the system is ``frozen''. We choose $t^*$ large enough to ensure that 
the system is in the hydrodynamic limit but not too large, so that region $A$ contains a finite 
density of unbound particles. As we will discuss in section~\ref{sec:conf}, the latter condition is 
crucial to observe bound-state confinement. 

As anticipated, in both protocols $(a)$ and $(b)$ of Fig.~\ref{fig0:quench}, since $|N_p\rangle$ is not an eigenstate of~\eqref{eq:ham}, the part $A$ of the chain starts to ``melt'' from arbitrarily small times $t=\epsilon>0$. 
First, the dynamics of an infinite homogeneous chain prepared in $|N_p\rangle$ gives rise to a Generalized Gibbs Ensemble~\cite{calabrese2016introduction} 
($GGE$) with extensive thermodynamic entropy. Provided that $\ell$ is large enough, local properties in the bulk of 
region $A$ are  described by the $GGE$ arising in the homogeneous chain. Oppositely, at the interface between region $A$ and the 
rest, the dynamics leads to the formation of a nontrivial profile. 
Specifically,  in the long-time limit ballistic transport sets in. As it will be explained in Section \ref{sec:conf}, provided that $t\ll \ell$ to ensure that the central region has 
a finite density of excitations, $GHD$ description is valid. This means that expectation values of 
generic local observables ${\mathcal O}(x,t)$ (with $x$ measured from the interface between $A$ and 
rest) are functions of $\zeta:=x/t$, and not of space 
and time separately. The expectation values of ${\mathcal O}(\zeta)$ can be obtained (see section~\ref{sec:ghd}) 
by using the Generalized Hydrodynamics~\cite{castro-alvaredo2016emergent,bertini2016transport} ($GHD$). 
Crucially, since the $XXZ$ chain is interacting, for sufficiently large $\Delta>1$ transport of 
bound states (of arbitrary size) of elementary excitations from region $A$ to $B$ is inhibited in the ballistic regime. This 
means that the expectation values ${\mathcal O}(\zeta)$ for $\zeta>0$ are fully determined by magnonic 
excitations. This bound-state confinement happens for both the protocols of 
Fig.~\ref{fig0:quench} at times $t\ll \ell$. As already stressed, for the protocol in Fig.~\ref{fig0:quench} $(a)$ 
the dynamics at asymptotically long times leads to zero 
density of magnonic excitations in the bulk of $A$. As we will discuss in section~\ref{sec:conf} this will lead 
to the deconfinement of bound states.

\section{Thermodynamic Bethe Ansatz ($TBA$) for quenches from homogeneous states in the $XXZ$ chain}
\label{sec:tba}

To derive bound-state confinement after the trap expansion of the $p$-N\'eel states (cf.~\eqref{eq:pneel}), 
it is necessary to construct the $GGE$ describing the steady state after the quench from the \emph{homogeneous} 
$p$-N\'eel states. A self-contained discussion of the derivation of the $GGE$ for generic prequench initial states is 
reported in~\ref{app:tba_gge}.

Here, we employ the framework of the Thermodynamic Bethe Ansatz~\cite{takahashi1999thermodynamics} ($TBA$). 
The eigenstates of the $XXZ$ chain are identified by a set of complex numbers $\lambda_j$, which 
are called rapidities. Each eigenstate is written as $|\{\lambda_j\}\rangle$, and the rapidities 
$\lambda_j$ label the different quasiparticles in the system. In the thermodynamic limit 
$L\to\infty$ at fixed density of quasiparticles, it is convenient to work with the rapidity density. 
Moreover, in the thermodynamic limit generic integrable systems exhibit different \emph{families} of 
quasiparticles. They correspond to bound states of elementary, magnon-like, excitations. 
The corresponding rapidities form string patterns in the complex plane~\cite{bethe1931zur}. 
Then, any 
thermodynamic macrostate of the system is characterized by a set of rapidity density $\{\rho_j(\lambda)\}_{j=1}^{\mathcal N}$, 
where ${\mathcal N}$ depends on the model under consideration and where, for $j>1$, 
$\rho_j=\rho_j(\mathrm{Re}(\lambda))$ with $\mathrm{Re}(\lambda)$ the center of the string. From now on, $\lambda$  
will denote the string center, and we will drop the real part in 
$\mathrm{Re}(\lambda)$. For the $XXZ$ chain at $\Delta>1$ one 
has ${\mathcal N}\to\infty$, meaning that bound states of arbitrary size are present. Similar to the 
densities $\rho_j(\lambda)$, one defines the hole densities $\rho_{h,j}(\lambda)$. The particle and hole 
densities for any legitimate $TBA$ macrostate satisfy the $TBA$ equation~\cite{takahashi1999thermodynamics} as 
\begin{equation}
	\label{eq:rho-decoup}
	\rho_{t,n}(\mu)=a_n(\mu)-a_{n,j}\star \rho_j, 
\end{equation}
where $\rho_{t,n}=\rho_n+\rho_{h,n}$ is the total density. The $\star$ symbol in~\eqref{eq:rho-decoup} 
denotes the convolution, i.e., 
\begin{equation}
	\label{eq:conv}
	(f\star g)(\mu):=\int_{-\pi/2}^{\pi/2}d\lambda f(\mu-\lambda)g(\lambda). 
\end{equation}
In~\eqref{eq:rho-decoup}, we defined $a_n$ as~\cite{takahashi1999thermodynamics}  
\begin{equation}
	\label{eq:a-def}
	a_n(\mu)=\frac{1}{2\pi}\frac{\sinh(n\eta)}{\cosh(\eta)-\cos(2\mu)}, 
\end{equation}
where the parameter $\eta$ is related to the anisotropy $\Delta$  as 
\begin{equation}
	\label{eq:eta}
	\eta:=\mathrm{arccosh}(\Delta). 
\end{equation}
Moreover, the parameters $a_{n,m}$ in~\eqref{eq:rho-decoup} are obtained  as 
\begin{equation}
	\label{eq:dTheta}
	a_{n,m}(\lambda):=\partial_\lambda\Theta_{n,m}(\lambda)=(1-\delta_{nm})a_{|n-m|}(\lambda)+2 
	a_{|n-m|+2}(\lambda)+\cdots 
	+a_{n+m-2}(\lambda)+a_{n+m}(\lambda), 
\end{equation}
where $a_n(\lambda)$ is defined in~\eqref{eq:a-def}. 
Here $\Theta_{n,m}$ is related to the scattering phase 
$S_{n,m}(\lambda)$ between a $n$-particle  and  a $m$-particle bound state as 
$\Theta_{n,m}(\lambda)=-i\ln(S_{n,m})$. 
For the following, it is useful to notice  that 
the ``bare'' velocities of the quasiparticles are given as 
\begin{equation}
	\label{eq:v-bare}
	v^{b}_{n}=\frac{\partial \varepsilon_n(\mu)}{\partial\mu}\left(\frac{\partial p_n}{\partial\mu}\right)^{-1}, 
\end{equation}
where the bare energy $\varepsilon^b_n$ and the quasimomentum $p^b_n$ are defined as 
\begin{equation}
	\varepsilon^b_n(\mu)=\frac{\sinh(\eta)\sinh(n\eta)}{\cosh(n\eta)-\cos(2\mu)},
	\quad p^b_n(\mu)=-i\sum_{j=1}^n\ln\left[\frac{\sin\left(\mu+i\frac{\eta}{2}(n+1-2j)+i\frac{\eta}{2}\right)}
	{\sin\left(\mu+i\frac{\eta}{2}(n+1-2j)-i\frac{\eta}{2}\right)}\right]. 
\label{eq:energy and mom def}
\end{equation}
Notice that Eq.~\eqref{eq:rho-decoup} allows to obtain only one of the 
$TBA$ densities, either $\rho_n(\lambda)$ or $\rho_{h,n}(\lambda)$.  
Crucially, to uniquely identify a thermodynamic macrostate, one  has  to 
determine both the particle and the hole densities. 
The information about the expectation values of local and 
quasilocal conserved quantities over the initial state allows to fix 
the $TBA$ densities identifying the $GGE$ that describes the steady state 
after the quench \cite{ilievski2016string,ilievski2016quasilocal}. 
In particular, the hole density is fixed by the initial-state 
expectation values of quasilocal charges $X_n(\lambda)$ by the relationship 
\begin{equation}
	\label{eq:X-rhoh_main}
	\rho_{h,n}(\lambda)=a_n-X_n\left(\lambda+i\frac{\eta}{2}\right)-X_n\left(\lambda-i\frac{\eta}{2}\right), 
\end{equation}
where 
$a_n$ is the same as in ~\eqref{eq:a-def} and $\eta$ is defined in ~\eqref{eq:eta}. An explicit expression for $X_n$ \eqref{eq:trace-X-1} is given in \ref{app:tba_gge}.
Hence, Eq.~\eqref{eq:rho-decoup} and Eq.~\eqref{eq:X-rhoh_main} form a system of equations that can be solved to univocally determine 
$\rho_n(\lambda)$ and $\rho_{h,n}(\lambda)$.

In the following, instead of $\rho_n(\lambda)$ and $\rho_{h,n}(\lambda)$, we will 
often consider the ratios $\eta_n(\lambda)$ and $\vartheta_n(\lambda)$ defined as 
\begin{equation}
	\label{eq:filling}
	\eta_n:=\frac{\rho_{h,n}}{\rho_{n}},\quad \vartheta_{n}(\lambda):=\frac{\rho_n(\lambda)}{\rho_{t,n}(\lambda)}. 
\end{equation}
Finally, the expectation value of any local observable $\mathcal{O}$ on a 
generic thermodynamic macrostate $GGE$ is computed from 
the quasiparticle densities $\rho_n(\lambda)$ as~\cite{takahashi1999thermodynamics} 
\begin{equation}
	\langle\mathcal{O}\rangle_{GGE}=\sum_n\int d\lambda\rho_n(\lambda)o_n(\lambda), 
\end{equation}
where $o_n(\lambda)$ depends on the observable under consideration. For instance, for the 
density of quasiparticles one has $o_n(\lambda)=n$.

\section{Generalized Hydrodynamics ($GHD$) for quenches from inhomogeneous initial states}
\label{sec:ghd}

Here we introduce the framework of Generalized Hydrodynamics~\cite{castro-alvaredo2016emergent,bertini2016transport} 
($GHD$) to describe the expansion dynamics (see Fig.~\ref{fig0:quench}).
Generalized Hydrodynamics is an extension of standard hydrodynamics to 
integrable systems. 
Integrable systems possess an extensive number of conserved charges. A crucial consequence 
of this property is that the many-body scattering is factorized into two-body elastic processes. 
Scattering processes preserve the bound states and the rapidities. 
This means that quasiparticles (i.e., magnons and bound states of magnons) are 
stable. Also the fact that the total number of  quasiparticles for any given value of the rapidity 
is preserved suggests that the spatial densities of quasiparticles satisfy a continuity equation, 
which is the main equation of $GHD$. 

In the standard $GHD$ setup, two semi-infinite homogeneous but different initial states are joined together 
at $t=0$ at the origin $x=0$. At $t>0$ the full system evolves under a homogeneous integrable Hamiltonian. 
This is the so-called bipartite quench protocol~\cite{alba2021generalized}. 
Notice that although the setup of the bipartite quench is different from the one described in Fig.~\ref{fig0:quench} 
(a), the two protocols give the same dynamics provided that $\ell$ (see Fig.~\ref{fig0:quench}) 
is large enough to ensure that the bulk of region $A$ is in the hydrodynamic limit and 
$1\ll v_\mathrm{max}t\ll\ell$ is large, but short enough to neglect the fact that region $A$ is finite. 
Here $v_\mathrm{max}$ is the maximum velocity in the system, i.e, $v_\mathrm{max}=\max_{n,\zeta,\lambda} v_{n,\zeta}(\lambda)$. 
Under these conditions the dynamics around the two edges of region $A$ happen independently and coincide 
with the dynamics in the bipartite protocol. 

For typical initial 
states in the bipartite setup, at long times and large 
distances  with the ratio $\zeta:=x/t$ fixed, the behavior of the system is described by a $\zeta$-dependent 
$GGE$. This $GGE$ is fully characterized by a set of $TBA$ densities 
$\rho_{n,\zeta}(\lambda)$ and $\rho_{h,n,\zeta}(\lambda)$. 
The $TBA$ densities satisfy the $TBA$ equations~\eqref{eq:rho-decoup} for any $\zeta$ and the $GHD$ equation, 
which  reads as~\cite{bertini2016transport,castro-alvaredo2016emergent,alba2021generalized} 
\begin{equation}
	\label{eq:ghd}
	[\zeta-v_{n,\zeta}(\lambda)]\partial_{\zeta}\vartheta_{n,\zeta}(\lambda)=0, \quad\mathrm{with}\, \zeta:=x/t. 
\end{equation}
Here $\vartheta_{n,\zeta}$ are the filling functions (cf.~\eqref{eq:filling}) defined as 
\begin{equation}
	\label{eq:filling}
	\vartheta_{n,\zeta}(\lambda)=\frac{\rho_{n,\zeta}(\lambda)}{\rho_{t,n,\zeta}(\lambda)}, 
\end{equation}
with $\rho_{t,n,\zeta}=\rho_{n,\zeta}+\rho_{h,n,\zeta}$ the total density. 
In~\eqref{eq:ghd},  $v_{n,\zeta}$ are the dressed velocities, which are defined as  
\begin{equation}
	\label{eq:dressed-v}
	v_{n,\zeta}=\frac{[\partial_\lambda \varepsilon_{n,\zeta}]^{\mathrm{dr}}
	(\lambda)}{[\partial_\lambda p_{n,\zeta}]^{\mathrm{dr}}(\lambda)}, 
\end{equation}
where $\varepsilon_{n}(\lambda)$ and $p_{n}(\lambda)$  are respectively the bare energy and quasimomentum of the 
quasiparticles, defined in \eqref{eq:energy and mom def}.
Given a generic ``bare'' function of the rapidity $\tau^{\mathrm{b}}_{n}(\lambda)$, its dressed version 
$\tau_{n,\zeta}^\mathrm{dr}$ is obtained by solving the integral equation as~\cite{bonnes2014light,korepin1993quantum} 
\begin{equation}
	\label{eq:dressing}
	\tau^{\mathrm{dr}}_{n,\zeta}(\lambda)=\tau^{\mathrm{b}}_{n}(\lambda)+\sum_{m=1}^\infty
	\int_{-\frac{\pi}{2}}^{\frac{\pi}{2}} d\mu
	\partial_\lambda\Theta_{n,m}(\lambda-\mu)\vartheta_{m,\zeta}(\mu)\tau^{\mathrm{dr}}_{m,\zeta}(\mu), 
\end{equation}
where $\partial_\lambda \Theta_{n,m}(\lambda)$ is defined in~\eqref{eq:dTheta}, and the filling function 
$\vartheta_{n,\zeta}(\lambda)$ is given in~\eqref{eq:filling}. 
It is clear from the  definition of the momentum $p_{n}$ as~\cite{takahashi1999thermodynamics} 
\begin{equation}
	\label{eq:dressed-p}
	[\partial_\lambda p_{n,\zeta}(\lambda)]^{\rm dr}=2\pi\rho_{t,n,\zeta}(\lambda),
\end{equation}
that the dressing of $\partial_\lambda p_{n}$ is obtained by considering 
$\tau^{\mathrm{dr}}_{n,\zeta}=2\pi\rho_{t,n,\zeta}(\lambda)$ in~\eqref{eq:dressing}. 

To solve~\eqref{eq:ghd} one has to provide the boundary conditions at $\zeta=\pm\infty$. 
Let us focus on the bipartite protocol denoting by 
$\vartheta_{n,L}$ and $\vartheta_{n,R}$ the two thermodynamic macrostates, i.e., the $TBA$ filling functions, 
describing the steady state arising at infinite time after the quench from the left and right homogeneous states. 
The functions $\vartheta_{n,L},\vartheta_{n,R}$, provide the boundary conditions in~\eqref{eq:ghd}. 
Before proceeding, one can observe that Eq.~\eqref{eq:ghd} can be solved for each $\zeta$ separately. 
It is straightforward to check that the solution of~\eqref{eq:ghd} can be written as  
\begin{equation}
\label{eq:ghd-sol}
\vartheta_{n,\zeta}(\lambda)=\Theta_\mathrm{H}(v_{n,\zeta}-\zeta)[\vartheta_{n,L}(\lambda)
-\vartheta_{n,R}(\lambda)]+\vartheta_{n,R}(\lambda). 
\end{equation}
where $\Theta_\mathrm{H}(x)$ is the Heaviside step function. The dressed velocity $v_{n,\zeta}$ in~\eqref{eq:ghd-sol} 
is obtained by using~\eqref{eq:dressing} and~\eqref{eq:dressed-v}~\eqref{eq:dressed-p}. Notice that 
since $v_{n,\zeta}$ depends on $\vartheta_{n,\zeta}$ via~\eqref{eq:dressing}, Eq.~\eqref{eq:ghd-sol} is not 
an explicit solution of~\eqref{eq:ghd}. To obtain $\vartheta_{n,\zeta}$ one can use an iterative strategy, 
starting from~\eqref{eq:ghd-sol} fixing $v_{n,\zeta}=v^b_{n}$, with $v^b_{n}$ the ``bare'' velocities 
(cf.~\eqref{eq:v-bare}). Then, one substitutes Eq.~\eqref{eq:ghd-sol} in~\eqref{eq:dressing} to obtain the 
dressed $\partial_\lambda \varepsilon_{n,\zeta}$. By using the $TBA$ equation~\eqref{eq:rho-decoup} one obtains 
$\rho_{t,n,\zeta}$ and the new dressed velocities via~\eqref{eq:dressed-v}. Finally, these steps are iterated 
until convergence. 

In the following we focus on the macrostate with $\zeta=0$, which 
describes the interface between the two chains. We also consider the situation in which  
the right chain is prepared in the vacuum state, while the left chain is prepared in the same product state defined in the region A (see Fig.~\ref{fig0:quench}). This corresponds to 
$\vartheta_{n,R}=0$ for any $n$. 
Now, Eq.~\eqref{eq:ghd-sol} becomes  
\begin{equation}
	\label{eq:ghd-sol-1}
	\vartheta_{n,0}(\lambda)=\Theta_\mathrm{H}(v_{n,0})\vartheta_{n,L}(\lambda). 
\end{equation}
In the following we neglect the subscript $\zeta$ in $\vartheta_{n,\zeta}$ and $v_{n,\zeta}$  
since we always consider $\zeta=0$.

\section{Bound-state confinement after trap-expansion experiments}
\label{sec:conf}

Here we discuss the confinement of the bound states for the expansion protocol 
of Fig.~\ref{fig0:quench} for generic initial state of part $A$. 
The state of $A$  corresponds to  fixing $\vartheta_{n,L}$ 
in~\eqref{eq:ghd-sol-1}. Let us summarize our main results. 
We show that for moderately large $\Delta$ \emph{ballistic} transport of all the bound states is inhibited, for quite generic 
initial states of part $A$. Specifically, 
in the hydrodynamic limit only quasiparticles with $n=1$, i.e., unbound 
quasiparticles, are present at $\zeta\ge 0$. This means that all the bound states with $n>1$ remain confined 
within region $A$ in Fig.~\ref{fig0:quench} (a). 
	In the following, we will employ the results of section~\ref{sec:ghd}, which are valid for quenches from 
	bipartite initial states. Again, although our initial state is tripartite, for times $t\ll\ell$,  
	the bulk of $A$ is not affected by the presence of region $B$. 
	Thus, the dynamics around the edges of $A$, which is relevant to address confinement, 
	are exactly the same as the dynamics that one has in the bipartite protocol. As a consequence, 
	for $t\ll\ell$ and $|x|\ll\ell$ the state of the system is well approximated by a $\zeta$-dependent GGE 
	(see Section~\ref{sec:check-ghd} for some numerical checks). 
	At times $t\approx\ell$ 
	when information about the interface between $A$ and the rest reaches the bulk of $A$ one could still apply $GHD$ although 
	one should take into account that the bulk of $A$, and hence the filling functions $\vartheta_{n,L}$,   
	evolve with time.

Bound-state confinement can be understood perturbatively in the large 
$\Delta$ limit, and it relies on the behavior of the scattering phases~\eqref{eq:dTheta}. 
Indeed, generic initial states of region $A$ that give rise to 
$TBA$ macrostates with ${\mathcal O}(1)$ density of quasiparticles with $n=1$ will lead to 
confinement of all the bound states with $n\ge 2$. 
Although it is established perturbatively, bound-state confinement  survives up 
to $\Delta\approx 1$. Precisely, for the N\'eel state bound state confinement survives up to~\cite{alba2019towards} 
$\Delta=\Delta^*\approx 1.4$. Upon lowering $\Delta$ below $\Delta^*$, 
bound states with $n=2$ start to leave region $A$. 
For $\Delta\ll \Delta^*$ bound states with $n>2$ are not confined. However, while the threshold $\Delta^*$ for 
confinement of the two-particle bound states is understood perturbatively, this is not the case for the larger 
bound states. Indeed, the deconfinement of bound states with $n>2$ happens in the 
nonperturbative regime at $\Delta\approx 1$, and is sensitive to the details of the initial state. 

Bound-state confinement manifests itself in the fact 
that the dressed velocities $v_{n}(\lambda)$ at $\zeta=0$ are negative for 
any $\lambda$ for $n>1$. 
To proceed, let us assume that only  bound states 
with $n> n_a$, for some $n_a>1$, are confined. This implies that Eq.~\eqref{eq:dressing} 
for $\tau_{n}=\partial_\lambda \varepsilon_{n}$ becomes 
\begin{equation}
	\label{eq:dressing-v-3}
	\partial_\lambda \varepsilon^{\mathrm{dr}}_{n}(\lambda)=\partial_\lambda\varepsilon^{\mathrm{b}}_{n}(\lambda)-
	\sum_{m=1}^{n_\mathrm{a}}\int^{\frac{\pi}{2}}_{-\frac{\pi}{2}}\!\! \frac{d\mu}{2\pi}\partial_\lambda\Theta_{n,m}(\lambda-\mu)\vartheta_{m,L}(\mu)
	\Theta_\mathrm{H}(\partial_\mu\varepsilon_m^\mathrm{dr}(\mu))
	\partial_\mu\varepsilon_{m}^{\mathrm{dr}}(\mu), 
\end{equation}
where $n_\mathrm{a}$ is such that  $\partial_\lambda\varepsilon^\mathrm{dr}_{n}(\lambda)<0$ for any $\lambda$ and 
for any $n>n_\mathrm{a}$. Notice that in~\eqref{eq:dressing-v-3} we replaced $v_{n,0}$ with $\partial_\mu
\varepsilon_n^\mathrm{dr}$ in the argument of the Heaviside theta function. This is possible because $\rho_{t,n}(\lambda)$ (cf.~\eqref{eq:dressed-v}) 
is positive. Also, notice that in the integrand in~\eqref{eq:dressing-v-3} one has that $\partial_\lambda\Theta_{n,m}$ and 
$\vartheta_{n,L}$ are both positive functions. For~\eqref{eq:dressing-v-3} to be consistent with 
confinement of the bound states with $n>n_a$, we require that the solutions of~\eqref{eq:dressing-v-3} 
satisfy the condition 
\begin{equation}
	\label{eq:v-cond}
	\partial_\lambda\varepsilon_{n}^\mathrm{dr}(\lambda)<0,\forall n>n_\mathrm{a},\forall\lambda\,. 
\end{equation}
Clearly, while the dressing~\eqref{eq:dressing} for the $XXZ$ with $\Delta>1$ is a system 
of infinite equations since $n\in [1,\infty)$, the system~\eqref{eq:dressing-v-3} is finite. 
Still, given the solutions of~\eqref{eq:dressing-v-3}, one has to check the validity of~\eqref{eq:v-cond} 
for any $n>n_a$. 
Before proceeding, it is interesting to observe that not all the 
species of quasiparticles can be confined at the same time, at least in the hydrodynamic limit. 
Indeed,  if all the quasiparticles are confined Eq.~\eqref{eq:dressing-v-3} implies 
that $\partial_\lambda\varepsilon_n^\mathrm{dr}(\lambda)=\partial_\lambda\varepsilon_n^\mathrm{b}(\lambda)$. However,  
$\partial_\lambda\varepsilon_n^\mathrm{b}(\lambda)$ is not negative for all values of $\lambda$ (cf.~\eqref{eq:energy and mom def}), 
which is in contraddiction with the assumption that all the quasiparticles are confined. 
This implies that the suppression of ballistic transport observed for instance in the 
domain-wall quench~\cite{gobert-2005} cannot be attributed to the bound-state confinement.

The conditions under which  the systems~\eqref{eq:dressing-v-3} and~\eqref{eq:v-cond} admit 
a solution can be derived 
in the large $\Delta$ limit. First, let us observe that $0\le \vartheta_{n,L}(\lambda)\le 1$ for any $\lambda$.  
Moreover, in the large $\Delta$ limit, at the leading order in powers of $\Delta$, one has that 
$\partial_\lambda\Theta_{n,m}(\lambda-\mu)={\mathcal O}(1)$ for any $n,m$. 
Precisely, for large $\Delta$ one has that 
\begin{equation}
	\partial_\lambda\Theta_{n,m}(\lambda)\to\frac{1}{\pi}(n+m-1),\quad \mathrm{for}\,\, n,m\ne1, 
\end{equation}
and $\partial_\lambda\Theta_{n,m}(\lambda)\to(n+m)/\pi$ otherwise. 
Finally,  it is straightforward to check that the velocities 
$v_n^\mathrm{b}$ (cf.~\eqref{eq:v-bare}) at the leading order in the large $\Delta$ limit are 
\begin{equation}
	\label{eq:v-delta}
	v_n^\mathrm{b}(\lambda)=2z^{n-1}\sin(2\lambda)+{\mathcal O}(z^{n+1}),\quad \mathrm{with}\,\, z:=e^{-\eta}=\Delta-\sqrt{\Delta^2-1}\,. 
\end{equation}
Notice that at the leading order in $z$, $v_n^\mathrm{b}(\lambda)=\partial_\lambda \varepsilon_n^\mathrm{b}(\lambda)$, 
with $\varepsilon_n^\mathrm{b}(\lambda)$ 
the bare energy in~\eqref{eq:energy and mom def}. 
For large $\Delta$ one has $z\approx 1/(2\Delta)$. 
Moreover, unlike the dressed velocities, the bare ones do not depend on the pre-quench initial state, so~\eqref{eq:v-delta} 
holds irrespective of the initial state of $A$ (see Fig.~\ref{fig0:quench}). Crucially, 
while the bare velocities of the quasiparticles with $n=1$ are ${\mathcal O}(1)$, the bound-state velocities, in the regime $\Delta>1$, decay exponentially with the bound-state size $n$. 
This hierarchy in the bare velocities is responsible for the bound-state confinement. 

Let us consider the case with $\vartheta_{1,L}(\lambda)={\mathcal O}(1)$, i.e., the 
situation in which in region $A$ there is a finite density of unbound quasiparticles. This happens 
for typical states, such as the N\'eel state and the Majumdar-Ghosh state~\cite{alba2019towards}.  
Let us also assume that $n_\mathrm{a}=1$, i.e., bound states with $n\ge 2$ remain confined within $A$ during 
the expansion dynamics. 
Now, let us consider Eq.~\eqref{eq:dressing-v-3} with $n=2$. It is clear that 
the second term in the right-hand side dominates for large $\Delta$ because 
$\partial_\lambda\varepsilon_n^\mathrm{b}={\mathcal O}(z)$ for $n=2$, 
whereas the second term in~\eqref{eq:dressing-v-3} is ${\mathcal O}(1)$. 
Clearly, the same argument applies also for $n>2$, implying that all the bound states 
with $n\ge 2$ are confined for sufficiently large $\Delta$. 
Interestingly, as the only dependence on $\lambda$ in the second term in~\eqref{eq:dressing-v-3} is in 
$\Theta_{n,m}$, which becomes independent of $\lambda$ for large $\Delta$, the dressed velocities of the 
confined bound states   become ``flat'' as functions of $\lambda$. 
Upon lowering $\Delta$, the two terms in~\eqref{eq:dressing-v-3} become of the same order, and 
quasiparticles  with $n=2$ start to be transported in $B$ (see Fig.~\ref{fig0:quench}). 
Notice that it is difficult to characterize analytically 
the confinement scenario for $\Delta\to1$. Typically, we observe that there is a ``cascade'' effect. Precisely, upon 
lowering $\Delta$, transport of bound states with progressively larger $n$ become allowed (see Section~\ref{sec:ghd-num}). 

We now illustrate two situations in which the system can escape bound-state confinement. 
The key observation is that bound-state confinement relies on the fact that at any time 
during the expansion the bulk of $A$ is described by 
$\vartheta_{1,L}={\mathcal O}(1)$. 
This condition remains true at arbitrarily long times 
if region $A$ (cf. Fig.~\ref{fig0:quench}) is infinite. Oppositely, if region $A$ is finite, 
at long times $t\gg\ell$ the bulk of $A$ starts to be depleted and the filling functions $\vartheta_{n,L}$ 
vary with time.  One could still apply~\eqref{eq:dressing-v-3} after taking into account the time dependence in 
$\vartheta_{n,L}$. The main effect is that  the second term in~\eqref{eq:dressing-v-3} 
becomes smaller. Indeed, one should observe that in Eq.~\eqref{eq:dressing-v-3}, 
	$\vartheta_{1,L}\Theta_H(\partial_\mu \varepsilon^\mathrm{dr}_1(\mu))$ accounts  
	for the scattering with the magnons arriving at $\zeta=0$ from the bulk of $A$. 
	Now, the fact that at long times  
	the bulk of $A$ cannot supply the magnons, because they left $A$, 
	suggests that 
	$\vartheta_{1,L}$  should 
	be small. The reasoning above and Eq.~\eqref{eq:dressing-v-3} 
	hold irrespective of the precise dynamics in the bulk $A$. 
	Notice that in principle one could apply 
	the $GHD$ directly to the tripartite protocol, following, for instance, Ref.~\cite{bulchandani2018bethe}. 
The reasoning above allows us to conclude that  for $t\gg \ell$  the bound states 
with $n=2$ can leave region $A$ after the ``evaporation'' of the unbound particles.

As for the second scenario in which there is no bound-state confinement, we observe that 
the condition $\vartheta_{1,L}={\mathcal O}(1)$ is not valid for generic initial states of $A$. 
For instance, it is violated  by  all the  $p$-N\'eel states with generic $p>1$. 
By using the framework of section~\ref{sec:tba}, 
one can check that $\vartheta_{n,L}={\mathcal O}(z^2)$ 
for $n<p$, whereas $\vartheta_{n,L}={\mathcal O}(1)$ for $n\ge p$. 
In turn, by solving the $TBA$ equations~\eqref{eq:rho-decoup} for the particle densities $\rho_n(\lambda)$, 
this implies that $\rho_p={\mathcal{O}(1)}$, whereas $\rho_{n\ne p}$ are suppressed in the large $\Delta$ limit. This 
reflects that since the  $p$-N\'eel state contains $p$ down spins on consecutive sites it has overlap mostly with 
eigenstates of the $XXZ$ chain that contain bound states of $p$ magnons.  
This has striking consequences on transport properties and bound-state confinement. 
Indeed, we anticipate that the fact that $\vartheta_{n,L}={\mathcal O}(z^2)$ implies a strong suppression of bound-state 
transport for $p\ge 4$. Let us first consider the case with $p=2$. Now, Eq.~\eqref{eq:dressing-v-3} suggests that 
bound states with $n=2$ are not confined. Indeed, the term with $m=1$ in the right-hand-side of~\eqref{eq:dressing-v-3} 
is ${\mathcal O}(z^2)$ and it is negligible compared to $\partial_\lambda\varepsilon_2^\mathrm{b}={\mathcal O}(z)$. 
This means that unlike the case with $p=1$ the two-particle bound states cannot be confined by the magnons, whereas all the 
bound states with $n>2$ are confined. We observe a similar behavior for $p=3$ (see section~\ref{sec:ghd-num}), i.e., that the bound states with $n\le p$ are 
not confined, whereas the ones with $n>p$ are all confined. 
Now, a different scenario emerges for $p\ge 4$. Indeed, for $p=4$ we observe that since $\partial_\lambda\varepsilon_4^{\mathrm{b}}={\mathcal O}(z^3)$, it is 
clear that the second term in~\eqref{eq:dressing-v-3} dominates at large $\Delta$ because of the scatterings with the quasiparticles with $n\le4$, 
which contribute with terms with smaller powers of $z$. This means that for $p=4$ all the bound states with $n<p$ are not confined, 
whereas the ones with $n\ge p$ are confined. 
This implies that upon increasing $p$ the bound states that dominate 
the thermodynamic properties, i.e., the bound states with $n=p$, become confined, and transport is suppressed.

\subsection{Bound-state confinement: Numerical GHD results}
\label{sec:ghd-num}

\begin{figure*}[t]
\includegraphics[width=\linewidth]{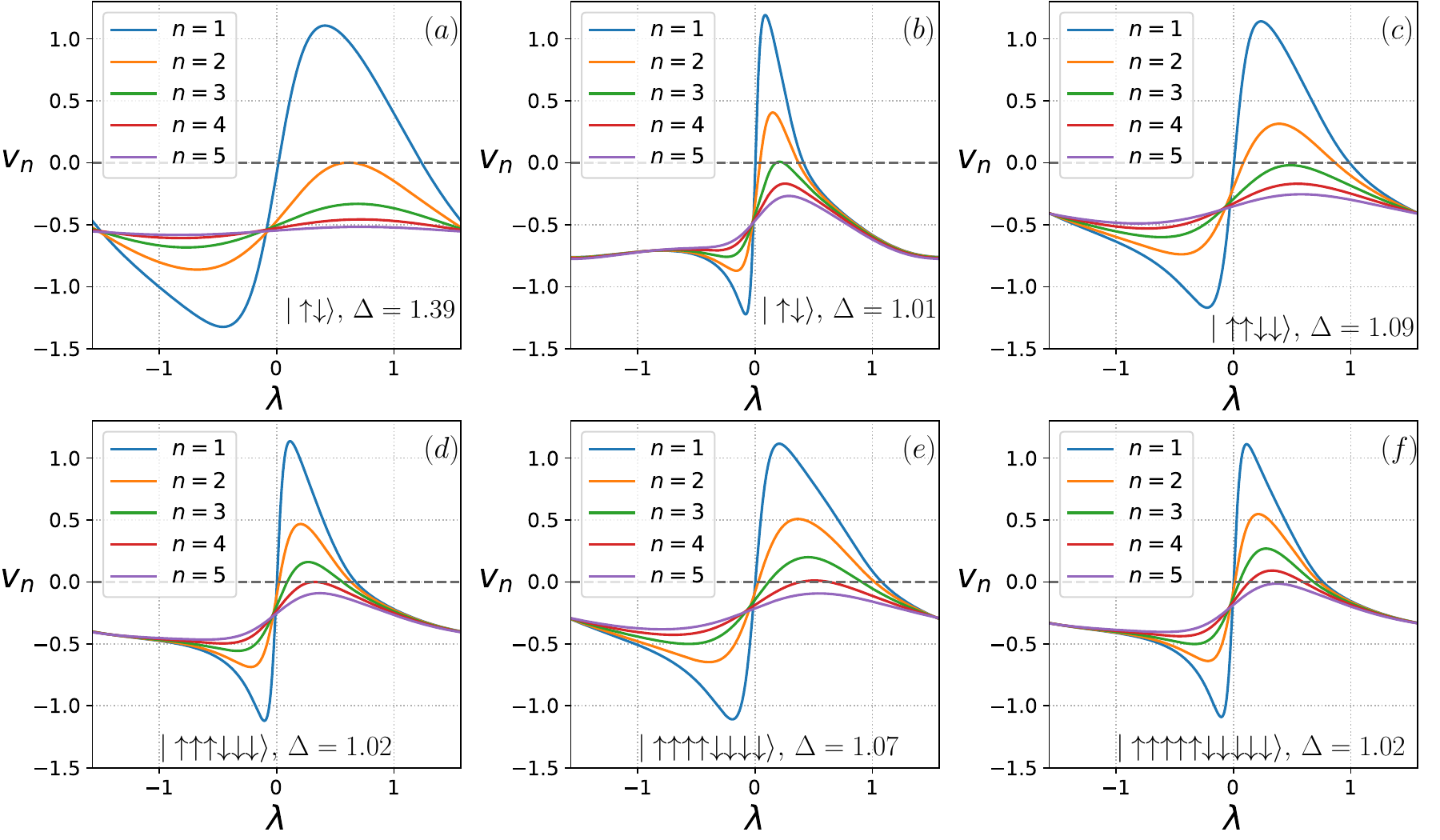}
\caption{ Group velocities $v_n$ of the $n$-particle bound states calculated over the 
 $GGE$ with $\zeta:=x/t=0$, which in the hydrodynamic limit describes  the 
 interface between region $A$ and the rest (Fig.~\ref{fig0:quench}). At $t=0$ 
 region $A$ is prepared in the $p$-N\'eel states (cf.~\eqref{eq:pneel}). 
 The different panels correspond to different values of 
 $p$ and $\Delta$.  Panel (a) and (b) are for $p=1$. Panels (c-f) show results for $p=2-5$. 
 In all the panels one has that $v_1(\lambda)$ is positive in an extended region in 
 $\lambda$, signaling that magnonic excitations are never confined. 
 (a) The case of $p=1$, i.e., the N\'eel state. At $\Delta\lesssim 1.4$ $v_2<0$ 
 for any $\lambda$, i.e., the two-particle bound states are confined. (b) Same as in 
 (a) for $\Delta=1.01$. While bound states with $n=2$ are not confined, 
 $v_3\le0$ for any $\lambda$, which means that three-particle bound states  are 
 confined. (c) The $p$-N\'eel state with $p=2$. Bound states with  
 $n\ge 3$ are confined. At $\Delta\approx 1.09$ three-particle bound states 
 are not confined anymore. (d) The case with $p=3$. At $\Delta\approx 1.02$ 
 the quasiparticles with $n=4$ are not confined. (e) The case with $p=4$. Now, even 
 quasiparticles with $n=4$ remain confined, at least up to $\Delta\approx 1.07$. 
 (f) The N\'eel state with $p=5$. Similar to (e), quasiparticles with $n=p$ are 
 confined. 
}
\label{fig:vel}
\end{figure*} 

Although qualitative features of bound-state confinement are understood perturbatively, to 
characterize quantitatively bound-state confinement it is necessary to determine numerically 
the dressed velocities $v_n(\lambda)$. 
Here we discuss numerical results confirming the scenario outlined in section~\ref{sec:conf}. 
Precisely, we numerically solve the $GHD$ equation~\eqref{eq:ghd} to obtain the 
dressed velocities $v_{n}$ for $\zeta=0$. The numerical solution of~\eqref{eq:ghd} is standard. 
By using a Gauss discretization, Eq.~\eqref{eq:ghd} becomes an infinite system of nonlinear equations, 
which can be truncated considering only bound states of size $n\le n_\mathrm{max}$. Finally, 
the truncated system can be solved with standard methods. 
By studying how the velocities change by varying $n_\mathrm{max}$, we 
observe that $n_\mathrm{max}=40$ is sufficient to have converged results.  

Our results are reported in Fig.~\ref{fig:vel}. We plot $v_n(\lambda)$ for $1\le n\le 5$ versus 
$\lambda$. We consider as initial states the $p$-N\'eel states (cf.~\eqref{eq:pneel}) with 
$1\le p\le5$. For $p=1$ the quench 
is the popular N\'eel quench. At large $\Delta$ one has that $v_n<0$ for any $\lambda$ and for 
any $n>1$, meaning that all the bound states are confined within region $A$. Fig.~\ref{fig:vel} 
(a) shows that for $\Delta\approx 1.4$ the bound states with less negative velocity $n=2$ start to be transmitted~\cite{alba2019towards}. 
Upon lowering $\Delta$, at $\Delta\approx 1.01$, the bound states with $n=3$ are not confined, 
as shown in Fig.~\ref{fig:vel} (b). 
For $p=2$, as anticipated in section~\ref{sec:conf}, the bound states with 
$n=2$ are not confined. 
%
\begin{figure}[t]
	\begin{center}
\includegraphics[width=.6\textwidth]{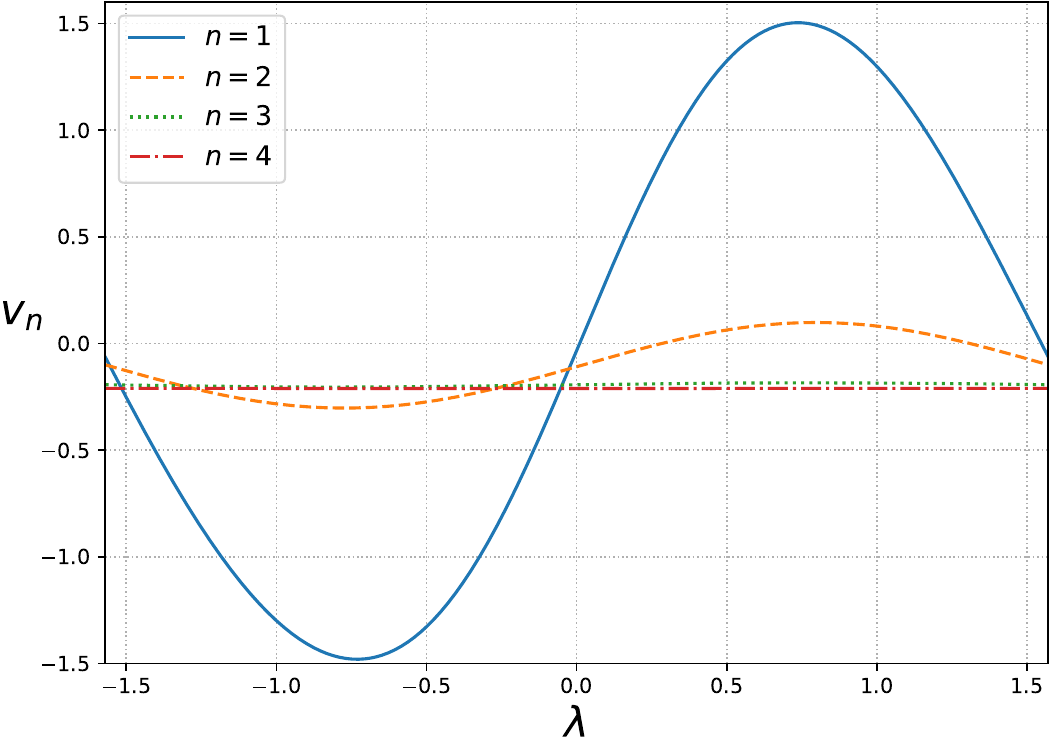}
\caption{ Group velocities $v_n$ of the $n$-particle bound states calculated over the 
 $GGE$ with $\zeta:=x/t=0$.  At $t=0$ 
 region $A$ is prepared in the $p$-N\'eel state (cf.~\eqref{eq:pneel}) with $p=2$. 
 The remaining part of the chain is prepared in the ferromagnet. 
 We show results for the $XXZ$ chain with $\Delta=5$. Notice that 
 the two-particle bound states are not confined because $v_2(\lambda)>0$ 
 in the region with $\lambda\approx 0.5$. On the other hand $v_3(\lambda)$ 
 and $v_4(\lambda)$ are negative for all values of $\lambda$, signaling 
 confinement. Moreover, both $v_3(\lambda)$ and $v_4(\lambda)$ exhibit a 
 ``flat'' behavior as functions of $\lambda$, as expected in 
 the large $\Delta$ limit. 
}
\label{fig:refe}
\end{center}
\end{figure} 
%
On the other hand, Fig.~\ref{fig:vel} (c) shows that for $p=2$ at $\Delta\approx 1.1$ the 
three-particle bound states are still confined in $A$. 
For $p=2$, we checked that absence of confinement for the two-particle bound states 
persists also at large $\Delta$.   
In Fig.~\ref{fig:refe} we report numerical results for the group 
velocities $v_n$ up to $n=4$ and the quench from the $p$-N\'eel state 
with $p=2$. The results are for $\Delta=5$. The fact that $v_2(\lambda)$ is positive in the region with $\lambda\approx 0.5$ 
signals that the bound states with $n=2$ are not confined. Moreover, we observe that larger bound states 
with $n=3,4$ are confined. The group velocities $v_3(\lambda)$ and $v_4(\lambda)$ are ``flat'', 
i.e., they exhibit a weak dependence on $\lambda$, in agreement with the discussion in section~\ref{sec:conf}.

A similar scenario occurs for 
$p=3$, i.e., the bound states with $n< 4$ are not confined, whereas the bound states with 
$n\ge4$ are all confined up to $\Delta\approx 1.02$. However, the scenario changes at $p=4$. As it is clear 
from Fig.~\ref{fig:vel} (e), for $p\ge 4$ the group velocities $v_p(\lambda)$ are negative 
for any $\lambda$. This  holds true for $p=5$ as well (see Fig.~\ref{fig:vel} (f)). 
Since the thermodynamic properties of the system are dominated by 
the bound states with $n=p$, this means that transport becomes more and more suppressed upon 
increasing $p$.

\section{Numerical tDMRG results}
\label{sec:numerics}

Here we discuss $tDMRG$ data~\cite{paeckel2019time} for the expansion protocols introduced in Fig.~\ref{fig0:quench}. 
Our results support the bound-state confinement scenario outlined in section~\ref{sec:conf}. 
As a preliminary check, in section~\ref{sec:check-ghd} we discuss $tDMRG$ data for the full-expansion 
protocol $(a)$ in Fig.~\ref{fig0:quench}, considering times $t\ll \ell$, which allows us to neglect the effects 
of the finite size of $A$ and use the $GHD$ results for the bipartite protocol. In our $tDMRG$ simulations we employ a Trotter 
decomposition of the evolution operator. During the dynamics the bond dimension $\chi$ of the Matrix Product 
State ($MPS$) representation of the wavefunction grows. At each time step we compress the $MPS$ 
by performing a Singular Value Decomposition, keeping the largest $\chi$  singular 
values. We increase the bond dimension $\chi$ until convergence of physical observables (see, for instance,~\ref{sec:bond}). 
Typically, in our simulations we use $\chi\approx 10^3$. 
Here we benchmark the results of section~\ref{sec:ghd} for 
several $p$-N\'eel initial states. 
In section~\ref{sec:full} we focus on the 
full-expansion protocol (see Fig.~\ref{fig0:quench} (a)) considering times $t\gg\ell$. 
In section~\ref{sec:cut-quench} we discuss the cut quench protocol (see Fig.~\ref{fig0:quench} (b)).  

\begin{figure*}[t]
\includegraphics[width=\linewidth]{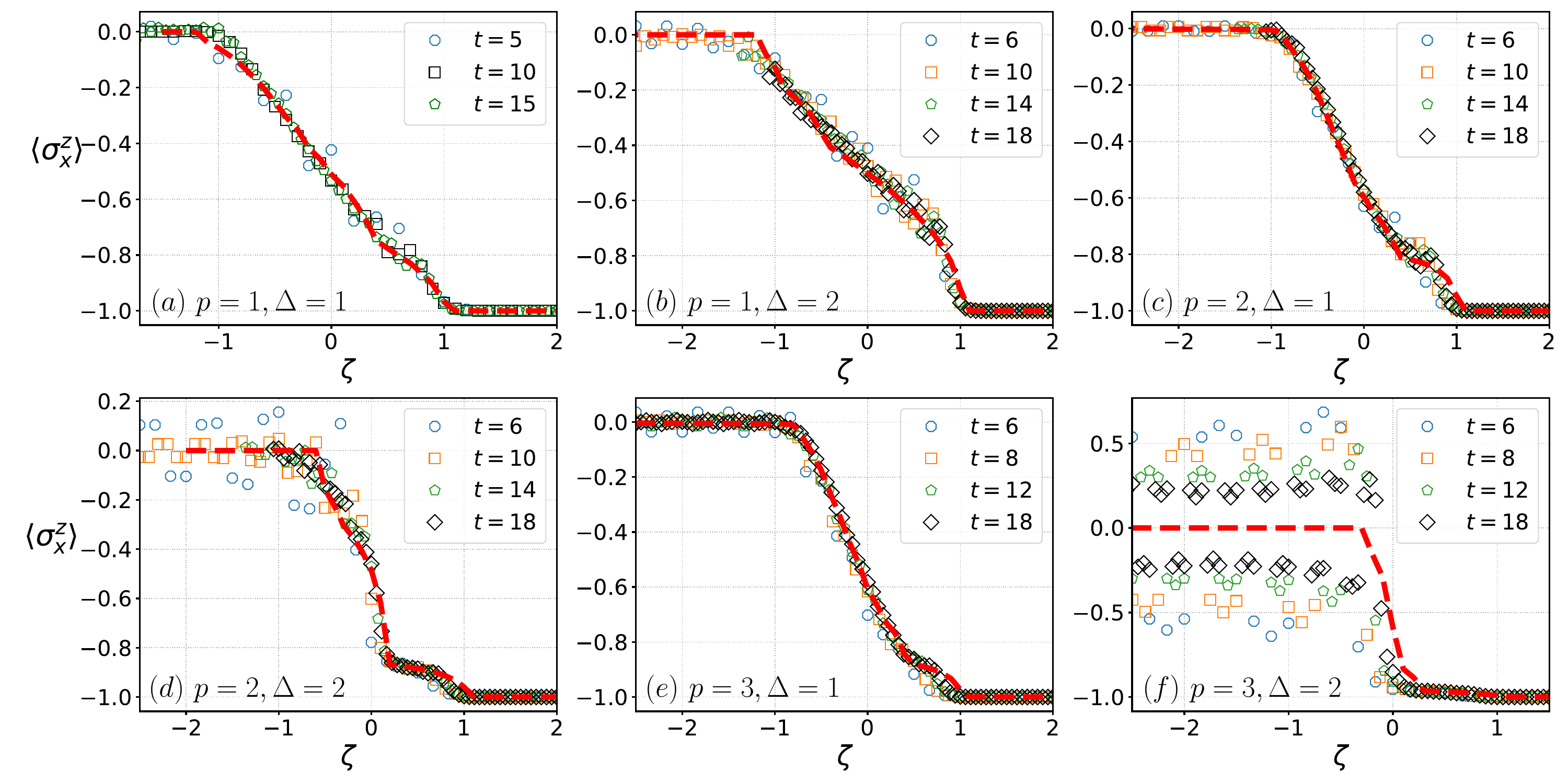}
\caption{Releasing the $p$-N\'eel states in the vacuum (see Fig.~\ref{fig0:quench}). 
 The figure shows $tDMRG$ data for the local magnetization $\langle\sigma_x^z\rangle$ 
 plotted as a function of $\zeta=x/t$. Here $x$ is 
 distance from the interface between region $A$ and the rest (Fig.~\ref{fig0:quench}). 
 The dashed line is the $GHD$ result valid in the limit $x,t\to\infty$ at fixed $\zeta$. 
 We show results for $p\le 3$ and $\Delta=1,2$. The different symbols denote different 
 times $t$. Notice that for $p=3$ and $\Delta=2$ 
 $t,x$ are not large enough to reach the hydrodynamic limit. 
}
\label{fig:scaling}
\end{figure*} 

\begin{figure}[t]
\begin{center}
\includegraphics[width=.7\linewidth]{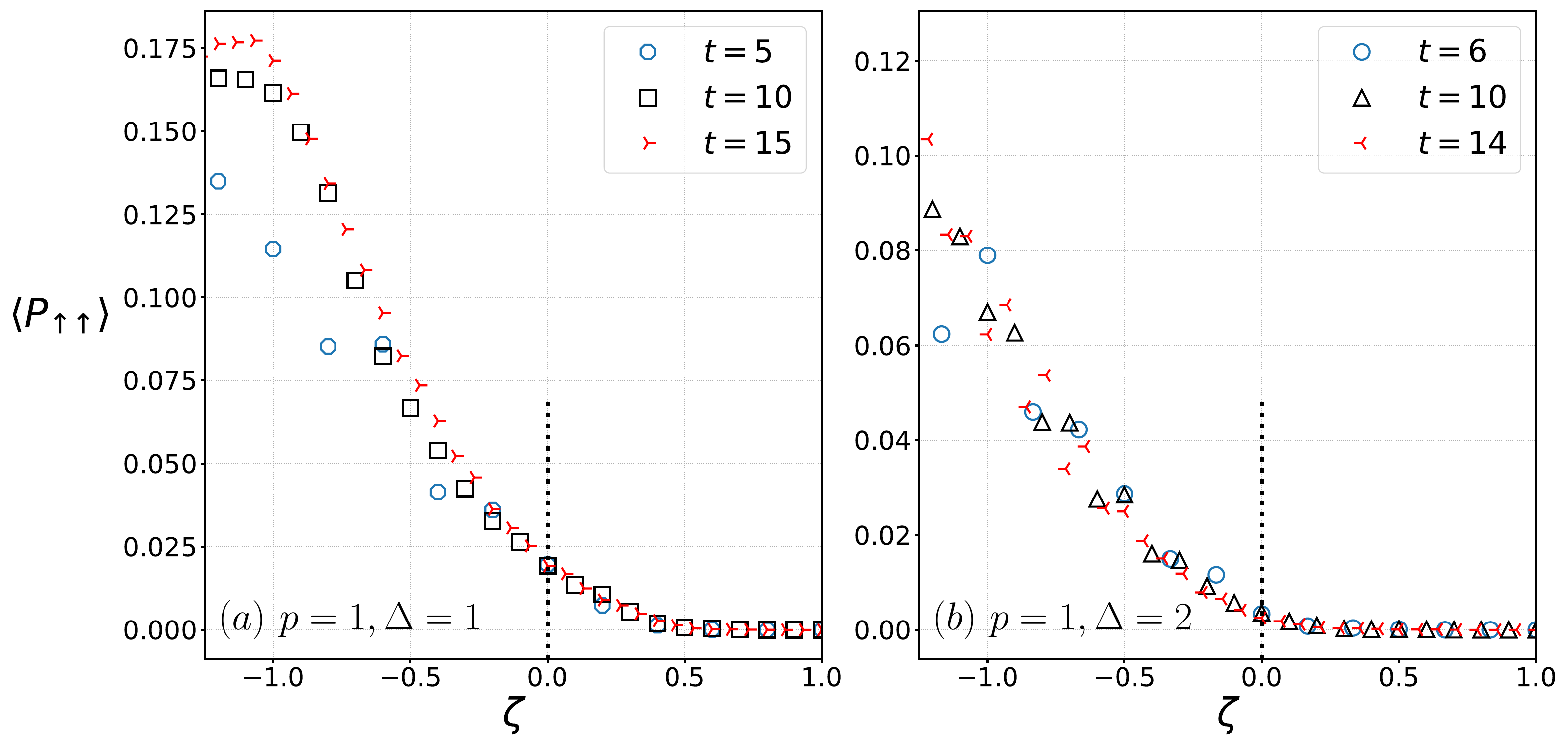}
\end{center}
\caption{Releasing the N\'eel states in the vacuum (see Fig.~\ref{fig0:quench}). 
 We plot the expectation value $\langle P_{x,\uparrow\uparrow}\rangle$ (cf.~\eqref{eq:project}) 
 versus the rescaled time $\zeta=x/t$, with $x$ the distance from the interface between 
 region $A$ and the rest (see Fig.~\ref{fig0:quench}). The size of region $A$ is $\ell=40$ sites. 
 Panels (a) and (b) show the expansion dynamics for $\Delta=1$ and $\Delta=2$, 
 respectively. The different symbols denote different times. 
 We only show results up to $t\lesssim 15$, to ensure that in the bulk of region $A$ 
 one has $\vartheta_{1,L}={\mathcal O}(1)$, which implies bound-state confinement (see section~\ref{sec:conf}). 
 Notice that at $\zeta=0$ for $\Delta=1$ one has a finite value for $\langle P_{x,\uparrow\uparrow}\rangle$, 
 whereas at $\Delta=2$ one has $\langle P_{x,\uparrow\uparrow}\rangle\approx0$, reflecting 
 bound-state confinement. 
}
\label{fig:P2collapse}
\end{figure} 

\subsection{Validity of GHD}
\label{sec:check-ghd}

Before investigating bound-state confinement, it is crucial to assess the validity of the 
$GHD$ description for the protocols discussed in Fig.~\ref{fig0:quench}. This is summarized in 
Fig.~\ref{fig:scaling}. We consider the expansion of the $p$-N\'eel states with $1\le p\le 3$. 
The different panels show $tDMRG$ data for the local magnetization $\sigma_x^z$ as a function 
of $\zeta=x/t$. For each $p$ we show results for $\Delta=1,2$. Here $x$ is the distance from the 
right edge of region $A$ (see Fig.~\ref{fig0:quench}). In our $tDMRG$ simulations we consider 
times $t\ll\ell$ to avoid effects due to the finite size of $A$. In Fig.~\ref{fig:scaling} 
we provide $tDMRG$ data for several times up to $t\approx 20$ (different symbols in the Figure). 
The $tDMRG$ data exhibit  collapse when plotted as function of $x/t$, confirming that $t\approx 20$ is large enough to 
ensure the hydrodynamic limit for all the values of $p$ and $\Delta$ considered, except for 
$p=3$ with $\Delta=2$ (see Fig.~\ref{fig:scaling} (f)). For $p=3$ and $\Delta=2$ the 
evolution in the bulk of region $A$ is slower compared to the other cases. 
Indeed, at $\zeta\le -1$, the magnetization profile exhibits strong oscillations, which are  
reminiscent of the structure of the initial state. In all the panels, the dashed curves are the 
$GHD$ results  obtained by solving numerically the $GHD$ equation~\eqref{eq:ghd}. The numerical results 
were obtained by truncating the infinite system of equations~\eqref{eq:ghd} keeping only 
the bound states with $n\le n_\mathrm{max}=40$. The $GHD$ equation~\eqref{eq:ghd} allows to 
obtain the occupation functions $\vartheta_{\zeta=0,n}(\lambda)$. The numerical results 
for $\vartheta_{0,n}(\lambda)$ can be substituted  in the $TBA$ equations~\eqref{eq:rho-decoup},  
after using that $\rho_{t,n}(\lambda)=\vartheta_{n}(\lambda)\rho_{n}(\lambda)$. 
Eq.~\eqref{eq:rho-decoup} allows one to obtain the quasiparticle densities $\rho_n(\lambda)$, 
which encode information about local observables. For instance, the local magnetization is 
given as 
\begin{equation}
	\langle\sigma_x^z\rangle=1-2\sum_{n=1}^\infty\int_{-\pi/2}^{\pi/2}d\lambda n \rho_n(\lambda). 
\end{equation}
For all the quenches that we explore in panels (a-e) of Fig.~\ref{fig:scaling} 
the $GHD$ results are in satisfactory agreement with the 
$tDMRG$ data, despite the limitation to times $t\lesssim 20$. For the expansion of the 
$p$-N\'eel state with $p=3$ and $\Delta=2$ (panel (f) in Fig.~\ref{fig:scaling}) deviations from 
the $GHD$ prediction are large. Also, notice that for $\zeta>0$ one has that $\langle\sigma_x^z\rangle\approx -1$, 
signaling that transport in general is strongly suppressed for $p=3$.

As anticipated in section~\ref{sec:conf}, for the $p$-N\'eel state with $p=1$, the 
bound states with $n=2$ remain confined in region $A$, as long as the system is in the hydrodynamic limit, and the 
$GHD$ description is valid. Specifically, this holds for $x,t\to\infty$ with $x/t$ fixed, with $x$ being 
the distance from the right edge of region $A$ (see Fig.~\ref{fig0:quench}). Moreover, 
for the bound-state confinement to happen, 
the condition $\vartheta_{1,L}={\mathcal O}(1)$ has to be satisfied. This is the case for 
$p=1$, at least if $t\ll\ell$, i.e., before the magnons leave region $A$.

Confinement of the two-particle bound states is discussed 
in Fig.~\ref{fig:P2collapse}. The Figure shows $P_{x,\uparrow\uparrow}$ (cf.~\eqref{eq:project}) 
plotted versus $\zeta=x/t$. The projectors $P_{x,\uparrow^{\otimes m}}$ are defined as 
\begin{equation}
	\label{eq:project}
	P_{x,\uparrow^{\otimes m}}:=\frac{1}{2^m}\prod_{j=0}^{m-1}(\mathds{1}+\sigma_{x+j}^z). 
\end{equation}
Although the projectors~\eqref{eq:project} are not \emph{bona fide} witnesses of 
bound states, we anticipate that $P_{x,\uparrow^{\otimes m}}$ are sensitive to the presence of a 
$m$-particle bound state at position $x$ \cite{morvan2022formation, ganahl2012observation}. 
Panels (a) and (b) show $tDRMG$ data for $P_{x,\uparrow\uparrow}$ for the $XXZ$ 
Hamiltonian with $\Delta=1$ and $\Delta=2$, respectively. In each panel the different symbols correspond 
to different times. For both $\Delta=1$ and $\Delta=2$ the $tDMRG$ data exhibit scaling as a function 
of  $\zeta$. Crucially, while for $\Delta=1$ (Fig.~\ref{fig:P2collapse} (a)) $\langle P_{x,\uparrow\uparrow}\rangle$ is nonzero 
for $\zeta>0$, one has $\langle P_{x,\uparrow\uparrow}\rangle\approx 0$ for any $\zeta>0$ for $\Delta=2$ (see Fig.~\ref{fig:P2collapse}). 
This reflects that while for $\Delta=1$ the bound states with $n=2$ are not confined, for $\Delta=2$ they 
remain confined in region $A$ during the expansion. Still, at $\zeta=0$, 
$\langle P_{x,\uparrow\uparrow}\rangle\approx 0.025$ for $\Delta=1$,  which suggests that the density of 
two-particle bound states that are not confined is ``small''. This means that  
it could be challenging to verify confinement, for instance experimentally.

\begin{figure*}[t]
\includegraphics[width=\linewidth]{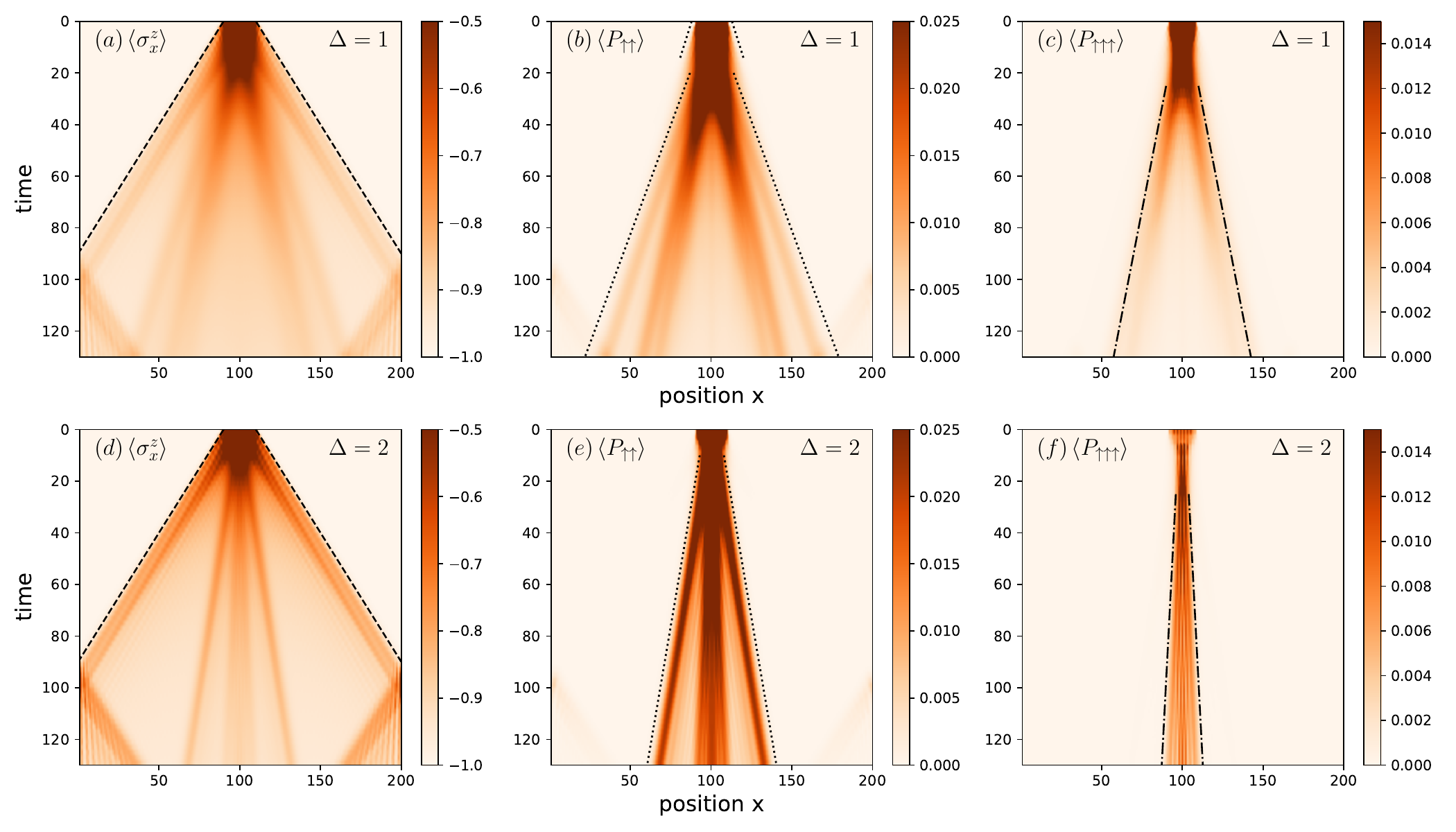}
\caption{ Releasing the N\'eel state in the vacuum. Region $A$ in Fig.~\ref{fig0:quench} of size $\ell=20$ 
 is prepared in the N\'eel state at $t=0$. The remaining part of the chain is in the 
 vacuum state, i.e., the ferromagnetic state. At $t>0$ the chain is let to evolve under the 
 $XXZ$ chain Hamiltonian respectively with $\Delta=1$ in the top row and $\Delta=2$ in bottom one. Panels (a,d) show $tDMRG$ data for the 
 local magnetization $\langle \sigma^z_x\rangle$ at generic position $x$ and at different times. 
 Notice the reflection at the boundary of the chain at long times. 
 Panels (b,e) and (c,f) show the projectors $P_{x,\uparrow\uparrow}$ and $P_{x,\uparrow\uparrow\uparrow}$.  
}
\label{fig:introD1D2}
\end{figure*} 

\subsection{Full expansion dynamics}
\label{sec:full}

In section~\ref{sec:check-ghd} we established the validity of the $GHD$ description for the quenches from the 
$p$-N\'eel states. Moreover, Fig.~\ref{fig:scaling} suggests that $t\approx 20$ is enough to access 
the hydrodynamic limit, which is essential to observe bound-state confinement. We now focus on 
the full expansion dynamics, i.e., considering times $t\gg\ell$.  We show $tDMRG$ data for the  expansion of the 
$p$-N\'eel states with $p=1$ for $\Delta=1$ and $\Delta=2$ in Fig.~\ref{fig:introD1D2}. The protocol is reported  in Fig.~\ref{fig0:quench}. Region $A$ at 
center of the chain is  prepared in the N\'eel state. Here we consider $\ell=20$. 
The reason for 
choosing $\ell=20$ is that it allows to simulate the dynamics using $tDMRG$ up to times 
$t\approx 100$. Indeed, entanglement, which is the major limitation of $tDMRG$, is produced in the bulk of region $A$ only, whereas the expansion does 
not contribute significantly to the entanglement growth. This also reflects that the $GHD$ dynamics preserves the 
thermodynamic entropy~\cite{alba2018entanglementand,alba2019entanglement}. As a consequence, the largest entanglement 
produced during the dynamics is bounded by $\ell$. 

The different panels in Fig.~\ref{fig:introD1D2} show different observables. 
Precisely, we consider the local magnetization $\sigma_x^z$, and the local projectors $P_{x,\uparrow\uparrow},P_{x,\uparrow\uparrow\uparrow}$ 
(cf.~\eqref{eq:project}). 
As anticipated, and has already observed in Ref.~\cite{ganahl2012observation} and Ref.~\cite{morvan2022formation}, 
$P_{x,\uparrow^{\otimes m}}$ is sensitive 
to bound states of $m$ spins starting from position $x$. 

Let us first discuss the dynamics with the $XXZ$ chain at $\Delta=1$. 
In Fig.~\ref{fig:introD1D2} (a,b,c) we plot the density of local magnetization $\langle\sigma_x^z\rangle$ as 
a function of position $x$ and time $t$. After the expansion starts, the central region 
``melts'' as quasiparticles are emitted at its edges. First, several ``jets'' of excitations are 
visible, traveling at different velocities. The fastest quasiparticles correspond to the 
magnonic excitations, i.e., with $n=1$, whereas the jets moving at smaller velocities correspond to 
bound states with $n=2$ and $n=3$ magnons. The dashed lines are the bare maximum 
group velocities of the quasiparticles (cf.~\eqref{eq:v-bare}). Since the initial state is expanded in the vacuum, 
the density of quasiparticles at the edges of the jets is small and interactions are ``weak'', which explains why 
the bare velocities describe the dynamics of the jets. In Fig.~\ref{fig:introD1D2} (b) we show 
the dynamics of the projector $P_{x,\uparrow\uparrow}$. Now, the outermost jet in 
Fig.~\ref{fig:introD1D2} (a) is not present, confirming that it is due to the quasiparticles with 
$n=1$. Oppositely, the two innermost jets of Fig.~\ref{fig:introD1D2} (a), corresponding to $n=2,3$ are still present. 
Similar behavior was observed in the expansion of few-spin initial states in Ref.~\cite{ganahl2012observation}. 
Importantly, one should observe that at short times $t\lesssim 20$, $P_{x,\uparrow\uparrow}$ exhibits a 
``halo'' feature at the edges of region $A$. This can be attributed to the fact that for $\Delta=1$ the 
two-particle bound states are not confined in the hydrodynamic limit. Notice, however, 
that the halo is quite weak, reflecting that the density of two-particle bound states  at the 
interface between $A$ and the rest is small. A similar scenario is observed for $P_{x,\uparrow\uparrow\uparrow}$ 
(see Fig.~\ref{fig:introD1D2} (c)). Now, the three-particle bound states behave as if they were confined 
up to $t\approx 20$. Notice that the fact (see Fig.~\ref{fig:vel} (b)) that $v_3\le 0$ at 
$\Delta\approx 1.01$ suggests that the density $\rho_3(\lambda)$ of three-particle bound states 
at the interface between $A$ and the vacuum could be small. This implies that even though bound states 
with $n=3$ are not confined at $\Delta=1$, this could be difficult to observe. 

Let us now discuss the same protocol in the $XXZ$ chain 
with $\Delta=2$ (see Fig.~\ref{fig:introD1D2}, bottom row). The qualitative behavior is the same 
as in the top row in Fig.~\ref{fig:introD1D2} . However, the projector $P_{x,\uparrow\uparrow}$ at short times $t\lesssim 20$ 
exhibits a ``bottleneck'' feature. In contrast with $\Delta=1$, at $\Delta=2$ the two-particle 
bound states are confined (see Fig.~\ref{fig:vel} (f)). Moreover,  in the hydrodynamic regime the density 
$\rho_2(\lambda)$ is zero for $\zeta=0$, and it is presumably small also at $\zeta\lesssim 0$. This 
explains the bottleneck in Fig.~\ref{fig:introD1D2} (d,e,f).
At long times after all the magnons have left region $A$, the two-particle bound states are allowed to leave 
region $A$, as discussed in section~\ref{sec:conf}, and as it is confirmed in Fig.~\ref{fig:introD1D2} (e). 
A similar dynamics happens for the three-particle bound states, as shown in Fig.~\ref{fig:introD1D2} (f).

\begin{figure}[t]
\begin{center}
\includegraphics[width=.45\linewidth]{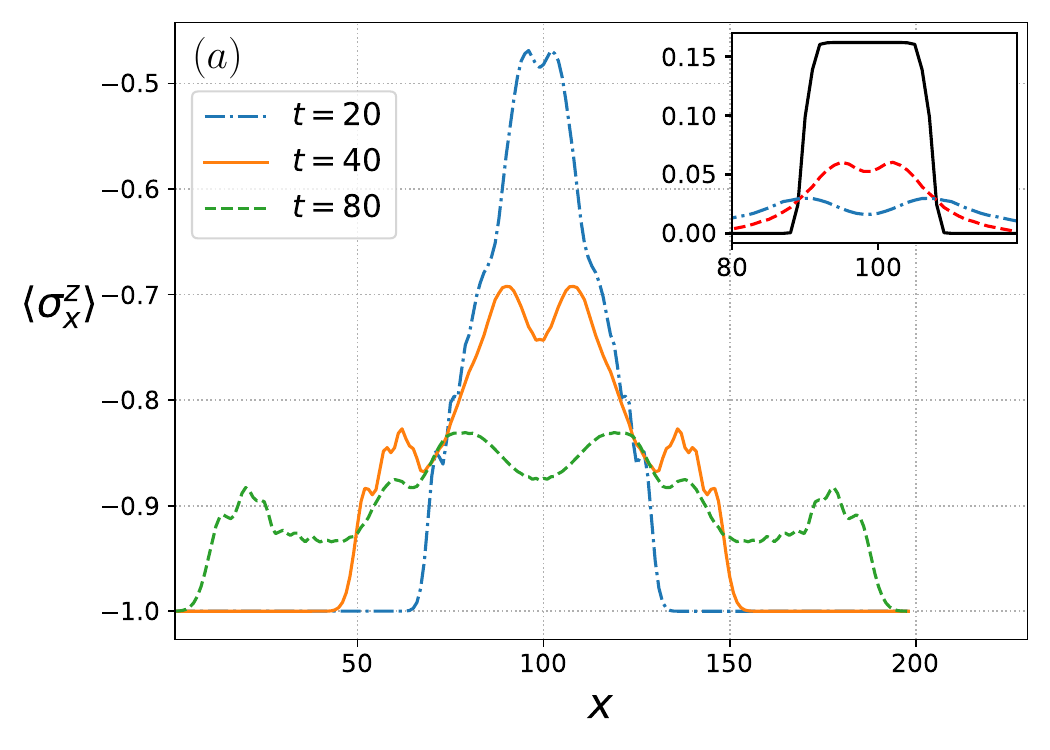}
\includegraphics[width=.45\linewidth]{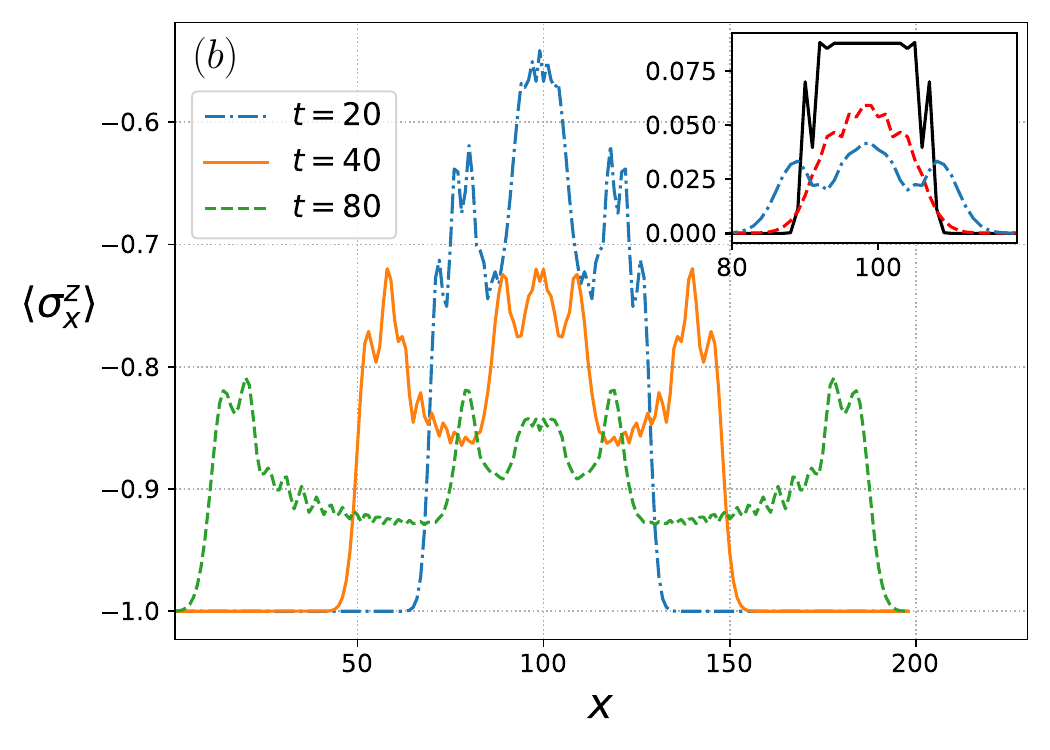}
\caption{ Releasing the N\'eel state in the vacuum. Same setup as in Fig.~\ref{fig:introD1D2}.
	Panels (a) and (b) show respectively the dynamics with $\Delta=2$ and $\Delta=1$.
 We show the magnetization profile $\langle \sigma_x^z\rangle$ as a function of the position 
 $x$ in the chain. The different curves correspond to different times. 
 More precisely, the blue (orange; green) curve shows the magnetization profile at time $t=20$ ($t=40$; $t=80$).
 Notice the different peaks, which correspond to different families of excitations. Precisely, the outermost peaks  
 correspond to magnonic excitations, whereas the innermost  ones to the  bound states with $n\ge 2$. 
}
\label{fig:profiles}
\end{center}
\end{figure} 

Let us now discuss the space-time profiles of local observables. 
In Fig.~\ref{fig:profiles} (a) and (b) we show the profile of the local magnetization $\langle\sigma_x^z\rangle$ 
after the expansion of the N\'eel state for the $XXZ$ chain with $\Delta=1$ and $\Delta=2$, respectively. 
Both panels in Fig.~\ref{fig:profiles} confirm that for $\Delta=1$ and $\Delta=2$ the bound states 
with $n=2$ and $n=3$ are not confined at long times, as already observed in Fig.~\ref{fig:introD1D2}. 
The $tDMRG$ data for $\Delta=1$ (left panel in Fig.~\ref{fig:profiles}) show a rapid melting of the profile. 
As the dynamics proceeds, $\langle\sigma_x^z\rangle$ exhibits a multi-peak 
structure, each peak corresponding to a different bound-state family. For $\Delta=1$ the peaks are 
blended together, reflecting that the bound states of different types start to leave region $A$  
at the same time. In particular, already at $t=40$ the bound states 
with $n=2$ are moving towards the edges of region $A$. In the inset we show the dynamics 
of $P_{x,\uparrow\uparrow}$, which is sensitive to quasiparticles with $n\ge 2$ only. The inset 
shows a rapid melting of $P_{x,\uparrow\uparrow}$. In contrast, for $\Delta=2$ (right panel in Fig.~\ref{fig:profiles}) 
the dynamics is slower, and the peaks corresponding to different $n$ are well separated. Notice that 
at $t=80$ the lump around $x=100$ corresponds to the three-particle bound states, which remain almost 
locked at the center of the chain. This observation suggests that for $\Delta=2$ the trap-expansion protocol 
can be used to ``distil'' states with large bound-state content. The inset shows the dynamics of $P_{x,\uparrow\uparrow}$. 
Notice that for $t\lesssim 10$ the bound states with $n=2$ are still within region $A$. 
Importantly, for $\ell=20$ the density of quasiparticles with $n=1$ in the bulk of $A$  starts to be small 
already at $t\approx 10$, implying that bound-state confinement is  not supported anymore (see section~\ref{sec:conf}). This behavior 
does not  violate GHD analytical prediction. In fact, for times $t\gtrsim20$, the bipartite GHD formulation is not anymore  predictive for the tripartite protocol due to the finite length of the region $A$.

\begin{figure*}[t]
	\begin{center}
\includegraphics[width=\linewidth]{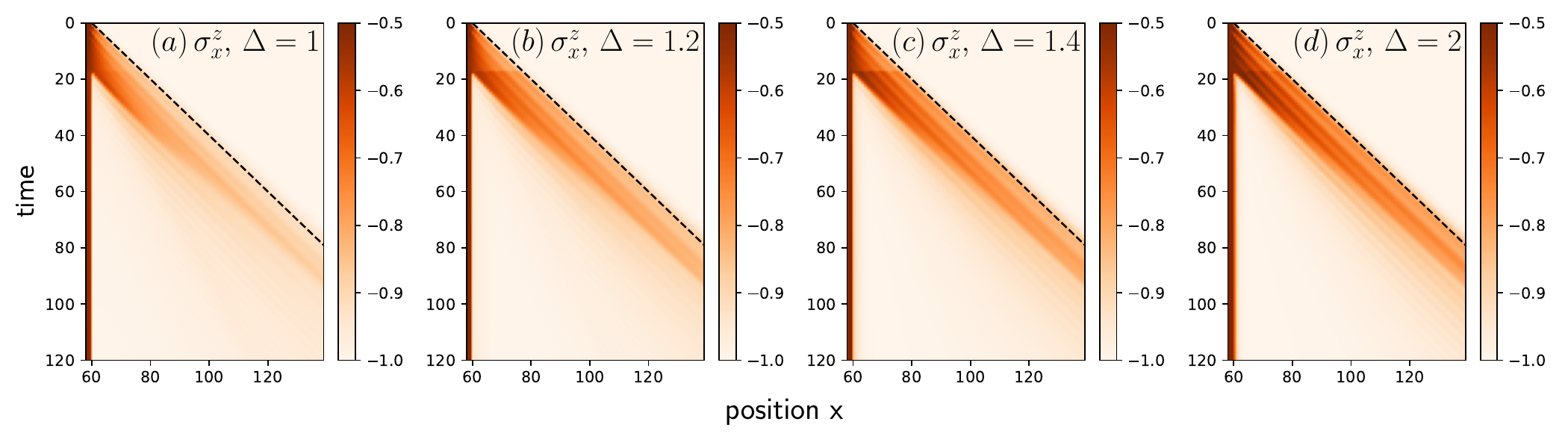}
\includegraphics[width=.5\linewidth]{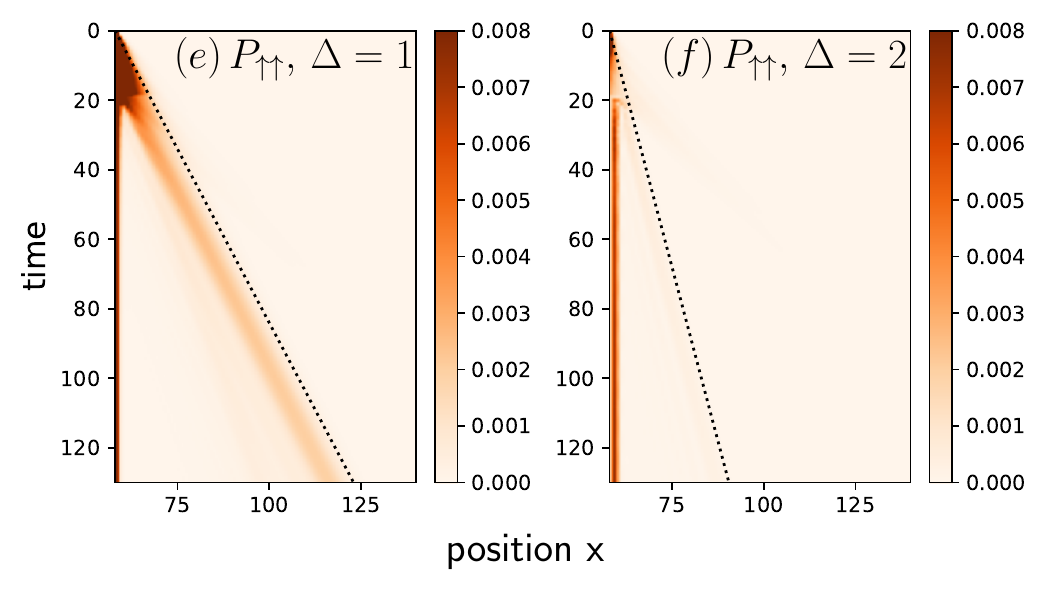}
\end{center}
\caption{ Releasing the N\'eel state in the vacuum. At $t=0$ the central part $A$ (see Fig.~\ref{fig0:quench}) 
 of a spin chain containing $\ell=20$ sites is prepared in the N\'eel state. At $t>0$ the 
 chain is let to evolve under the $XXZ$ chain Hamiltonian with different values of $\Delta=\{1, 1.2, 1.4, 2\}$. For $0<t\le 20$ the 
 full chain evolves. At $t>20$ only the $B$ part (see Fig.~\ref{fig0:quench}) is evolved. 
 Panels (a)-(d) show $tDMRG$ data for the local magnetization $\langle\sigma_x^z\rangle$ as a function 
 of the position $x$ in the chain. Panels (e)-(f) show results for the projector $P_{x,\uparrow\uparrow}$. 
 Notice that at $\Delta=2$, $P_{x,\uparrow\uparrow}$ vanishes, signalling bound-state 
 confinement. The sudden jumps around $t\approx 20$ are numerical artifacts due to the fact that we truncate 
 the $MPS$ representation of the system for $t>t^*$. 
}
\label{fig:neelP1P2}
\end{figure*} 

\subsection{Dynamics after the cut protocol}
\label{sec:cut-quench}

To investigate  bound-state confinement, it is also convenient to employ the protocol in 
Fig.~\ref{fig0:quench} (b).
The full chain is now evolved with the $XXZ$ Hamiltonian up to 
a time $t^*$. Here $t^*$ is chosen large enough to ensure the validity of the hydrodynamic 
limit. As it is clear from section~\ref{sec:check-ghd}, $t\approx 20$ is sufficient to 
ensure the hydrodynamic limit. Moreover, $t^*$ is chosen short enough compared to $\ell$ to have 
$\vartheta_{1,L}={\mathcal O}(1)$ in the bulk of $A$, which is essential to have bound-state 
confinement. Here we choose $t^*=20$ and $\ell=40$. 
We consider the $p$-N\'eel states with $p=1-3$. 

The cut protocol allows  to accurately study the nature  of the excitation that at time 
$t>t^*$ are in the region on the right of $A$. In fact, all the quasiparticles 
that are confined in region $A$ for $0<t<t^*$, are "frozen" by the cut protocol 
and do not contribute to the dynamics.

\subsubsection{$p$-N\'eel state with $p=1$}
\label{sec:cut-p1}

Our results for the N\'eel state (i.e., for $p=1$) are shown 
in Fig.~\ref{fig:neelP1P2}. Panels $(a-d)$ show the local magnetization $\sigma_x^z$, as a function 
of position $x$ and time $t$. In the Figure we show $tDMRG$ data. The $tDMRG$ algorithm is the same as 
in section~\ref{sec:check-ghd} and~\ref{sec:full}. In our simulations we keep the bond dimension $\chi\approx 1000$ 
for $t\le t^*$. We observe that typically for $x>0$, i.e., in region $B$, the bond dimension needed to 
accurately describe local observables is significantly smaller as compared with $x<0$ (see~\ref{sec:bond}). 
This is expected since entanglement is produced in region $A$ only. 
Thus, since at $t>t^*$ we evolve only region $B$ we further reduce the bond dimension to $\chi\approx 800$. 
As it is clear from the Figure, for $t>t^*$ (see Fig.~\ref{fig0:quench}), under the evolution with 
the Hamiltonian~\eqref{eq:ham} 
restricted to $x>0$ the quasiparticles emitted from $A$ propagate ballistically in the vacuum. 
The dependence on $\Delta$ is weak. In the bottom row in Fig.~\ref{fig:neelP1P2} (panels (e-f) in 
the Figure) we report the dynamics of $P_{x,\uparrow\uparrow}$. For $\Delta=1$, 
$\langle P_{x,\uparrow\uparrow}\rangle$ 
exhibits a ballistic spreading. The nonzero signal reflects that the bound states with $n=2$ are not 
confined. $P_{x,\uparrow\uparrow}$ becomes smaller 
upon increasing $\Delta$. For instance, it is barely visible at $\Delta=2$. Notice that in the 
hydrodynamic limit one should expect that $\langle P_{x,\uparrow\uparrow}\rangle$ is exactly zero 
for $\Delta>\Delta^*\approx 1.4$ (see Fig.~\ref{fig:vel}).

\subsubsection{$p$-N\'eel state with $p=2$}
\label{sec:cut-p2}

\begin{figure}[t]
\begin{center}
\includegraphics[width=\linewidth]{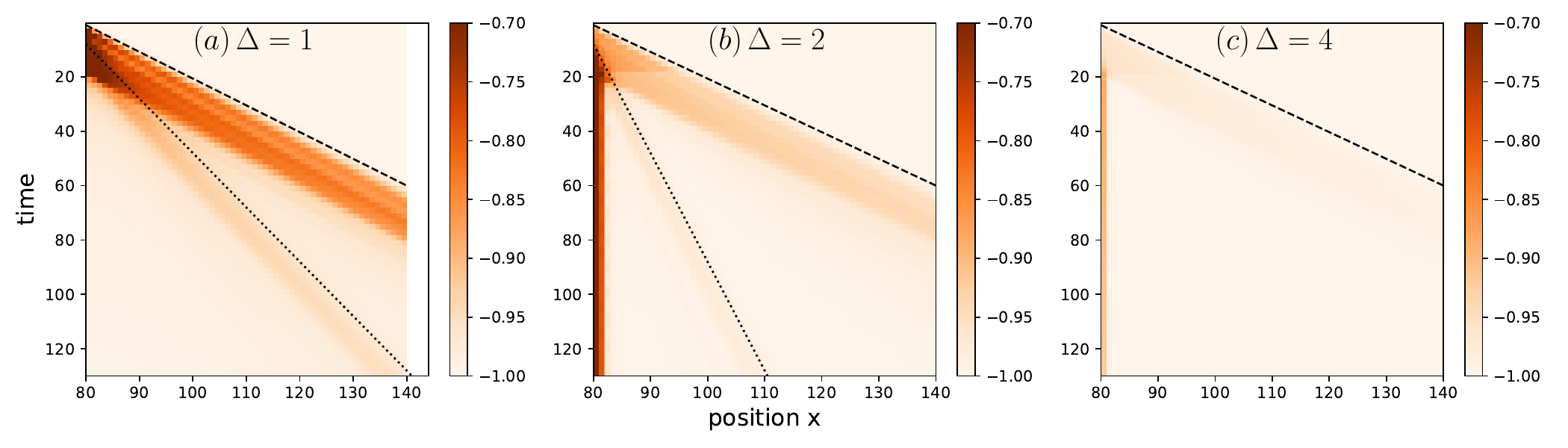}
\includegraphics[width=.5\linewidth]{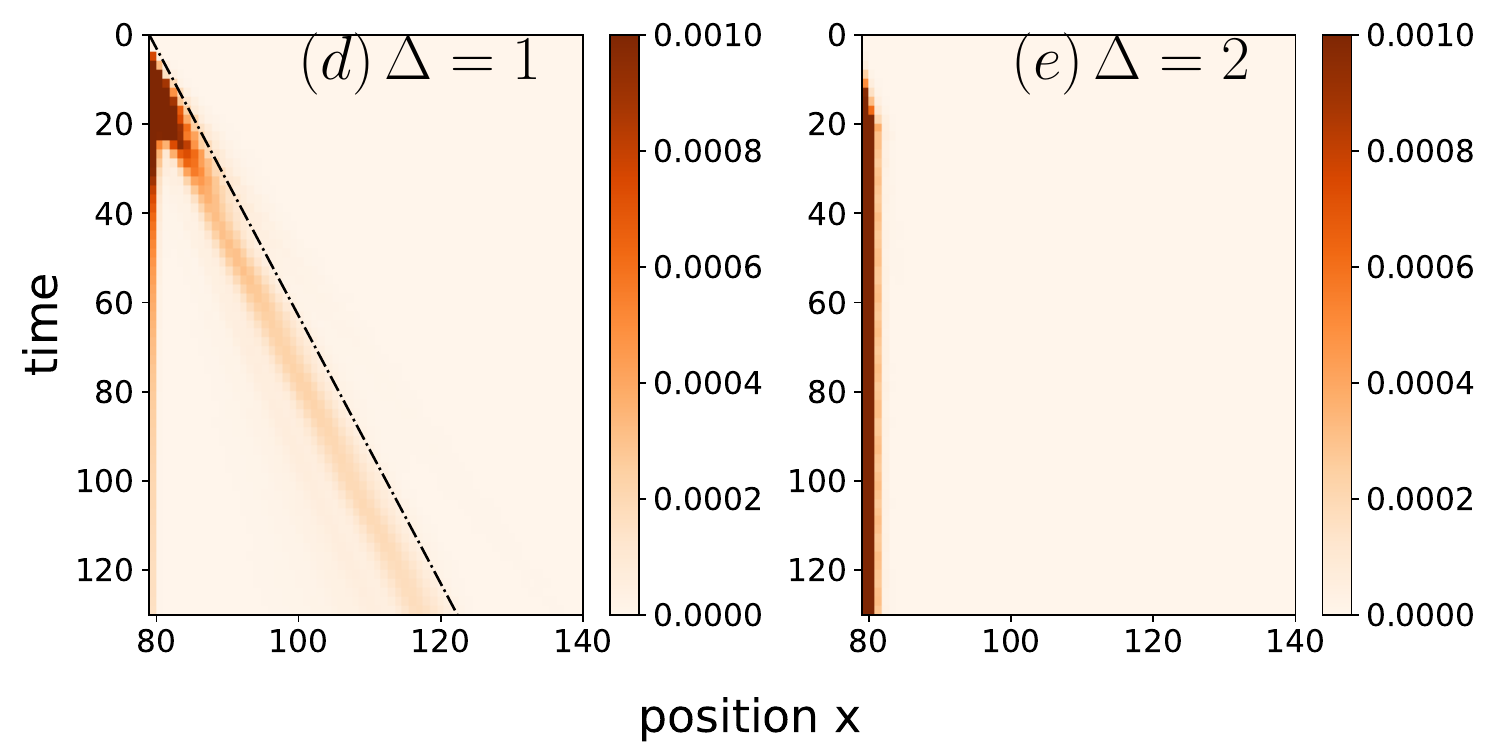}
\end{center}
\caption{ Releasing the ``fattened'' $p$-N\'eel state with $p=2$ in the vacuum in the cut protocol (see Fig.~\ref{fig0:quench} (b)). 
 The figure shows $tDMRG$ data for the local magnetization $\langle\sigma_x^z\rangle$ as a function 
 of $x$ and the time $t$. The different panels (a-c) show results for different values of $\Delta$. 
 For $\Delta=1$ and $\Delta=2$ the two jets correspond to the magnonic excitations and the 
 two-particle bound states. 
 Panels (d,e) show  $\langle P_{x,\uparrow\uparrow\uparrow}\rangle$ as a function of 
 time and position $x$ for  $\Delta=1$ and $\Delta=2$. Notice that for $\Delta=2$ the quasiparticles with $n=3$  
 are confined in $A$. The origin of the sudden jumps visible at $t\approx 20$ in (a-c) is the same as explained in the caption of Fig. \ref{fig:neelP1P2}.
}
\label{fig:neel2P1}
\end{figure}

Let us now discuss the cut protocol (see Fig.~\ref{fig0:quench} (b)) for the $p$-N\'eel state with 
$p=2$. Our results are reported in Fig.~\ref{fig:neel2P1}. Panels (a)-(c) 
show $tDMRG$ data for the dynamics with the $XXZ$ chain Hamiltonian and $\Delta=1,2,4$. 
In the Figure we plot the local magnetization $\langle \sigma_{x}^z\rangle$ as a function 
of position $x$ and time $t$. For $\Delta=1$ (Fig.~\ref{fig:neel2P1} (a)) two distinct signals are 
visible. The two ``jets'' correspond to quasiparticles with $n=1$, i.e., free ones, and to the 
two-particle bound states. The two-particle jet is due to the fact that 
for $p=2$ the bound-state content of the chain is enhanced as compared to $p=1$, and 
two-particle bound states  are not confined at $\Delta=1$. Notice also that the bound states 
exhibit smaller group velocities as compared with the quasiparticles with $n=1$. The dashed lines 
reported in Fig.~\ref{fig:neel2P1} (a) are the maximum bare velocities 
$v_{n,\mathrm{max}}^\mathrm{b}=\max_\lambda(v_n^\mathrm{b}(\lambda))$ (cf.~\eqref{eq:v-bare}). 

In Fig.~\ref{fig:neel2P1} (b) we show data for $\Delta=2$. Crucially, the jet corresponding to 
the two-particle bound states is still visible. This is in agreement with the results of 
section~\ref{sec:conf}. Indeed, for $p=2$ 
one should expect that the quasiparticles with $n=2$ are never confined. On the other hand, 
as it is clear from Fig.~\ref{fig:vel} (c), the three-particle bound states are confined 
for $\Delta\gtrsim 1.1$. We should mention that, however, as the density of quasiparticles 
with $n=3$ is ``small'', it is numerically challenging to observe the three-particle bound states 
confinement. Finally, in Fig.~\ref{fig:neel2P1} (c) we show results for $\Delta=4$. Now, the 
jet corresponding to $n=1$ is weak, in contrast with the case with $p=1$. This is due to the fact 
that for $p=2$ the density $\rho_1(\lambda)$ even in the bulk of $A$ 
vanishes in the limit of $\Delta$ large, reflecting that thermodynamic properties of 
the system are dominated by the bound states with $n=2$. On the other hand, although one has $\rho_2(\lambda)=\mathcal{O}(1)$ 
in the bulk of $A$, the flux of two-particle bound states through the interface between $A$ and the rest is suppressed 
upon increasing $\Delta$ because the bound-state velocities, despite the dressing, decrease with $\Delta$. 
This explains  why the jet associated with the two-particle bound states is 
not visible in Fig.~\ref{fig:neel2P1} (c). Notice that this is a ``trivial'' suppression of 
transport, which depends in a continuous way on $\Delta$. In particular, the $1/\Delta$ dependence 
of $v_2^b$ would suggest a $1/\Delta$ suppression of two-particle bound states upon increasing 
$\Delta$. 
Finally, in Fig.~\ref{fig:neel2P1} (d,e) we report the projector $P_{x,\uparrow\uparrow\uparrow}$ 
after the expansion of the $p$-N\'eel state with $p=2$. In Fig.~\ref{fig:neel2P1} (d) and (e) 
we show results for $\Delta=1$ and $\Delta=2$, respectively. While for $\Delta=1$ ballistic spreading 
is visible, for $\Delta=2$, the projector $P_{x,\uparrow\uparrow\uparrow}$ is ``frozen''. 

\begin{figure}[t]
	\begin{center}
\includegraphics[width=.7\linewidth]{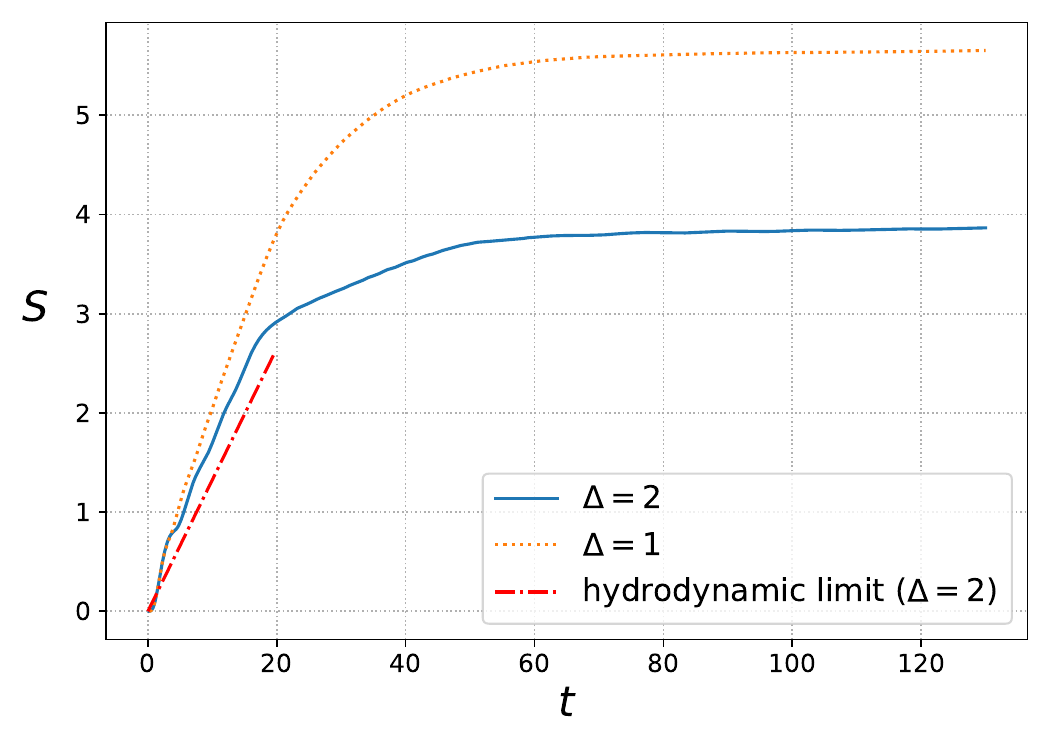}
\caption{ Releasing the N\'eel state in the vacuum. Dynamics of the 
 entanglement entropy. The curve shows $tDMRG$ data for a chain with 
 $L=200$ sites. The size of region $A$ (see Fig.~\ref{fig0:quench}) is 
 $\ell=20$. 
 The plot shows the entropy $S$ of subsystem $B$ (see Fig.~\ref{fig0:quench}). 
 Since subsystem $A$ includes 
 the N\'eel part, the growth of $S$ with time is due to the entanglement 
 between the excitations that leave the N\'eel part and those that are 
 left behind. At long times $S$ measures the entanglement between the 
 ``jets'' of excitations in Fig.~\ref{fig:introD1D2}. 
 Notice the multistage 
 growth of the entropy. At short times the growth is dominated by the 
 entangled magnonic excitations. The second slope visible in the 
 figure is attributed to the two-excitations bound states. The last 
 stage of the growth is due to the three-particle bound states.
}
\label{fig:entropy}
\end{center}
\end{figure} 

\section{Bound-state confinement and dynamics of the entanglement entropy}
\label{sec:ent}

Here we show that bound-state confinement is sharply reflected  in the 
dynamics of the von Neumann entropy. We focus on the full expansion of the N\'eel state  
(see Fig.~\ref{fig0:quench} (a)). 
Fig.~\ref{fig:entropy} shows the von Neumann entropy $S$ of region $B$ (see Fig.~\ref{fig0:quench} (a)). 
The dotted curve is the result for $\Delta=1$, 
whereas the full line shows the result for $\Delta=2$. First, for both values of $\Delta$ 
the von Neumann entropy saturates at long time. This reflects that the entanglement dynamics is 
due to the creation of pairs of entangled quasiparticles in region $A$~\cite{alba2017entanglement}. 
This is the content of the so-called quasiparticle picture~\cite{calabrese-2005,fagotti2008evolution,alba2017entanglement,alba2018entanglement}. 
The two quasiparticles forming the entangled pairs travel with opposite group 
velocities. As they propagate, they entangle larger regions. Precisely, 
the von Neumann entropy of a region is proportional to the number of 
entangled pairs that are shared between the region and the rest. The 
contribution of a shared pair characterized by rapidities $(-\lambda,\lambda)$ is 
the contribution of that pair to the thermodynamic entropy of the 
$GGE$  describing the steady state after the quench from the homogeneous N\'eel state. 
The quasiparticle picture holds in the hydrodynamic limit $\ell,t\to\infty$ with the 
ratio $t/\ell$ fixed. Notice that the saturation of the von Neumann entropy at long times 
is due to the finite-size of $A$. Indeed, since $A$ is finite, the number of 
entangled pairs that can generate entanglement between $A$ and $B$ is bounded by $\ell$. 
For a global quench from two semi-infinite chains prepared in nontrivial initial states, 
the entanglement entropy of each chain would grow indefinitely.

Let us focus on the results for  $\Delta=1$ in Fig.~\ref{fig:entropy}. 
The initial linear growth up to $t\approx 20$ is due to the entangled pairs that 
are initially produced in $A$ and at later times leave region $A$, and hence start to 
be shared with the rest (see Fig.~\ref{fig0:quench}). Again, the saturation of $S$ at long 
times reflects that region $A$ is finite. 
For $\Delta=1$, $S$ is essentially 
constant for $t\gtrsim 60$. The saturation of the entanglement entropy is the key to reach 
times long as $t\approx 120$ with $tDMRG$, because after the entropy saturates the computational cost 
is mild. One should also observe that the saturation of $S$ happens ``smoothly''. This is 
somewhat reminiscent of the fact that for $\Delta=1$ the bound states are not confined. 
Crucially, for $\Delta=2$, $S$ exhibits a  multistage linear growth before saturating. 
This reflects the bound-state confinement. Specifically, the linear growth up to $t\approx 20$ 
is due to the quasiparticles with $n=1$. In this early time regime, the slope of the linear growth 
should be the same as in the vacuum expansion of a semi-infinite chain prepared in the N\'eel state. 
The prediction of the quasiparticle picture is~\cite{alba2018entanglementand,alba2019entanglement} 
\begin{equation}
	\label{eq:slope-quasi}
	S=t \int d\lambda|v_{1,\zeta=0}(\lambda)|s^{YY}_{1,\zeta=0}(\lambda), 
\end{equation}
where $s_{1,\zeta=0}^{YY}$ is the Yang-Yang entropy~\cite{takahashi1999thermodynamics} of the macrostate with $\zeta=0$ that 
describes the interface between region $A$ and region $B$ (see Fig.~\ref{fig0:quench}). 
$s_{n,\zeta}^{YY}$ is readily obtained from the particle densities $\rho_{n,\zeta}(\lambda)$ 
 describing the $GGE$ as 
\begin{equation}
	\label{eq:yy}
	s_{n,\zeta}(\lambda):=\rho_{t,n,\zeta}\ln(\rho_{t,n,\zeta})-
	\rho_{n,\zeta}\ln(\rho_{n,\zeta})-\rho_{h,n,\zeta}\ln(\rho_{h,n,\zeta}),
\end{equation}
where the densities $\rho_{n,\zeta},\rho_{h,n,\zeta},\rho_{t,n,\zeta}$ are 
defined in section~\ref{sec:tba} (see also section~\ref{sec:ghd}). Notice that 
in~\eqref{eq:slope-quasi} only the quasiparticles with $n=1$ contribute, reflecting 
that all the bound states are confined in $A$, and cannot contribute to the 
von Neumann entropy. Now, the $tDMRG$ data in Fig.~\ref{fig:entropy} exhibit deviations 
from~\eqref{eq:slope-quasi}. This is expected because~\eqref{eq:slope-quasi} holds in 
the hydrodynamic limit $t\to\infty$. Finally, we should mention that it is possible to 
characterize the trajectories~\cite{alba2019entanglement} of the generic entangled pairs produced in $A$, which 
are responsible for the entanglement growth. This, in principle, could allow  to determine also 
the secondary slopes visible in Fig.~\ref{fig:entropy}. 

\section{Conclusions}
\label{sec:concl}

We investigated bound-state confinement after trap-expansion experiments in the $XXZ$ spin chain at $\Delta>1$. 
We showed that after the expansion in the vacuum of quite generic initial states the bound states of the 
$XXZ$ chain remain confined in the initial region, provided that $\Delta$ is large enough. Upon lowering $\Delta$ 
the bound states are allowed to leave region $A$, i.e., they are not confined anymore. The condition 
ensuring confinement is the presence of a finite density of magnonic excitations in the bulk of the initial region. 
Physically, bound-state confinement is a genuine consequence of interactions, i.e., the scattering, between the bound states and the magnons. 
Interestingly, if the size of the initial region is finite, at long times the density of magnons vanishes, as they 
leave the initial region, and bound-state confinement breaks down. Thus, for the expansion of finite regions,  
bound-state confinement is present only at intermediate times 
$t= {\mathcal O}(\ell)$. 

We investigated bound-state confinement after the expansion of the $p$-N\'eel states (see Fig.~\ref{fig0:quench}).  
In the limit $p\to\infty$ one recovers the domain wall initial state, and transport is suppressed. We showed that for 
small $p\lesssim 4$, bound-state confinement happens for the bound states of size $n>p$. At larger $p$, even the 
bound states with $n=p$ are confined, which is consistent with the absence of transport in the limit $p\to\infty$. 
Finally, we showed that bound-state confinement is reflected in a multi-stage linear growth of the entanglement entropy. 

Let us now discuss some directions for future research. First, we investigated bound-state confinement in 
a lattice model. It would be interesting to understand whether confinement occurs for systems in the continuum, 
such as the attractive Lieb-Liniger gas. This would pave the way for the experimental verification of bound-state 
confinement, for instance with atom-chip experiments~\cite{schemmer2019generalized,bouchoule2022generalized}. It would 
be also interesting to investigate confinement in lattice models that can be realized with  
trapped ions~\cite{kranzl2023observation} and cold-atom experiments~\cite{WeiDavid2022Qgmo}. Interestingly, bound-state confinement could be potentially 
used to prepare thermodynamic states with high bound-state content. Specifically, the protocol to enrich the bound 
state content would consists of two steps. In a first step one can expand the initial state as in Fig.~\ref{fig0:quench} (a). 
Then, after the magnonic excitation are ``evaporated'' one could trap the remaining excitations by applying an external 
potential. As the bound states are typically associated with lower thermodynamic entropy, bound-state confinement 
can also be used to prepare low-entropy states. It has been pointed out recently that bound-states 
could be robust against long-range interactions~\cite{macri2021bound,kranzl2023observation}. It would be interesting 
to understand whether long-range spin chains exhibit  bound-state confinement as well. Finally, it would be important 
to clarify whether bound-state confinement is also present in integrable quantum circuits~\cite{vanicat2018integrable,ljubotina2019ballistic,claeys2022correlations}.

\section*{Acknowledgements}

This study was carried out within the National Centre on HPC, Big Data and Quantum Computing - SPOKE 10 (Quantum Computing) and received funding from the European Union Next-GenerationEU - National Recovery and Resilience Plan (NRRP) – MISSION 4 COMPONENT 2, INVESTMENT N. 1.4 – CUP N. I53C22000690001. 
This work has been supported by the project ``Artificially devised many-body quantum dynamics in low dimensions - ManyQLowD'' funded by the MIUR Progetti di Ricerca 
di Rilevante Interesse Nazionale (PRIN) Bando 2022 - grant 2022R35ZBF.\\
This work has been partially funded by the ERC Starting Grant 101042293 (HEPIQ). 


\appendix
\section{$GGE$ thermodynamic macrostate in the $XXZ$ spin chain}
\label{app:tba_gge}
Here we provide a self-contained discussion of the derivation of the $GGE$ for a 
generic pre-quench initial state. We follow Ref.~\cite{ilievski2016string,ilievski2016quasilocal}. 

In particular, we show that the information about the expectation values of local and 
quasilocal conserved quantities over the pre-quench initial state allows to fix 
univocally the $TBA$ densities identifying the $GGE$ that describes the steady state 
after the quench. 

We start from the Algebraic Bethe Ansatz treatment of the $XXZ$ chain. 
A central object in the Algebraic Bethe Ansatz~\cite{korepin1993quantum} for the $XXZ$ chain 
is the spin-$j/2$ Lax operator $L_j(\mu)$ defined as 
\begin{equation}
	\label{eq:lax}
	L_j(\mu)=\frac{1}{\sinh(\eta)}\left(\begin{array}{cc}
		\sin(\mu+i\eta s^z) & i\sinh(\eta)s^-\\
		i\sinh(\eta)s^+ & \sin(\mu-i\eta s^z)
	\end{array}\right). 
\end{equation}
where $\eta$ is defined in~\eqref{eq:eta} and $\mu$ is a real parameter. 
For $j=1$, which corresponds to spin-$1/2$ representation, one has  
$s^\alpha=\sigma^\alpha/2$, with $\alpha=+,-,z$. Moreover,  $\sigma^+,\sigma^-$ are the standard 
combinations $\sigma^\pm:=\sigma^x\pm i\sigma^y$, with $\sigma^\alpha$ Pauli matrices. 
The Lax operator acts as a $2\times 2$ matrix over the physical Hilbert space $\mathcal{V}_{1}$ of the model.
The entries of~\eqref{eq:lax} live in the ``auxiliary'' Hilbert space $\mathcal{V}_j$ of dimension $j+1$. 
Thus, the global Hilbert space is $\mathcal{V}_{1}\otimes \mathcal{V}_j$. 
The auxiliary space is spanned by vectors $|n\rangle$ with $n=0,1,,\dots,j$, with $|0\rangle$ being 
a highest-weight vector. The operators $s^\alpha$ with $\alpha=\pm,z$ are $q$-deformed spin-$j$ operators, 
forming the $j$-representation of the $q$-deformed algebra $SU(2)$. They obey the 
commutation relations 
\begin{equation}
	[s^+,s^-]=[2s^z]_q,\quad 
	q^{\pm s^z}=q^{\pm 1}s^\pm q^{\pm s^z},\quad
	\mathrm{with}\,\, [x]_q:=\frac{\sinh(\eta x)}{\sinh(\eta)}. 
\end{equation}
The action of the operators $s^\pm,s^z$ on the states $|n\rangle$ is given as 
\begin{align}
	& s^z|n\rangle=\left(\frac{j}{2}-n\right)|n\rangle\\
	& s^+|n\rangle=\sqrt{[j-n]_q[n+1]_q}|n+1\rangle\\
	& s^-|n+1\rangle=\sqrt{[j-n]_q[n+1]_q}|n\rangle. 
\end{align}
Thus, the eigenvalues of $s^z$ are $-j/2,\dots,j/2$. 

As usual in Algebraic Bethe Ansatz~\cite{korepin1993quantum}, 
one can construct a set of conserved quantities from the transfer matrix $T_j(\mu)$, which is defined 
as 
\begin{equation}
	\label{eq:T}
	T_j(\mu)=\mathrm{Tr}_{\mathcal{V}_j}\left[L_j^{(1)}(\mu)L_j^{(2)}(\mu)\cdots L_j^{(L)}(\mu)\right],  
\end{equation}
where the trace is over the auxiliary space. 
In~\eqref{eq:T} the superscript $k$ in $L_j^{(k)}(\mu)$ is to stress that it acts in the 
space $\mathcal{V}_1\otimes \mathcal{V}_j$, where $\mathcal{V}_1$ is the spin-$1/2$ Hilbert space of 
site $k$ of the chain. Moreover, in~\eqref{eq:T}, $T_0(\mu)$ is given as 
\begin{equation}
	T_0(\mu)=\left(\frac{\sin(\mu)}{\sinh(\eta)}\right)^L\mathds{1}. 
\end{equation}
A complete set of local and quasilocal conserved charges $X_j(\mu)$, which commute with $H$, is 
defined as~\cite{ilievski2016quasilocal}  
\begin{equation}
	\label{eq:X}
	X_j(\mu)=\frac{1}{L}\frac{1}{2\pi i}\partial_\mu\ln\left(\frac{T^{[+]}_j(\mu))}{T^{[j+1]}_0(\mu)}\right), 
\end{equation}
where we introduced the notation 
\begin{equation}
	\label{eq:shift}
	F_j^{[\pm k]}(\mu) :=F_j\left(\mu\pm ik\frac{\eta}{2}\right), 
\end{equation}
for a generic operator $F_j(\mu)$, 
and $T_j(\mu)$ is defined in~\eqref{eq:T}. The transfer matrices $T_j(\mu)$ satisfy the 
commutation relations 
\begin{equation}
	[T_j(\mu),T_{j'}(\mu')]=0, \quad\forall j,j' \in\mathbb{Z}_{\ge 0},\,\forall \mu,\mu'\in \mathbb{C}. 
\end{equation}
One can also check that the transfer matrices $T_j$ in~\eqref{eq:tq} satisfy a 
Hirota equation (or the so-called $T$-system) as~\cite{ilievski2016quasilocal}  
\begin{equation}
	\label{eq:hirota}
	T_j^{[+]}T_j^{[-]}= T_0^{[j+1]}T_0^{[-j-1]}+T_{j-1}T_{j+1}. 
\end{equation}
Importantly, one can also verify the $T$-$Q$ relation as~\cite{bazhanov2010a}  
\begin{equation}
	\label{eq:tq}
	T_1 Q=T_0^{[+]}Q^{*[-2]}+T_0^{[-]}Q^{[2]}, 
\end{equation}
where $Q(\mu)$ is the so-called Baxter $Q$-operator.
The spectrum of the $Q$ operator is given as 	$Q(\mu):=\prod_{k=1}^M\sin(\mu-\mu_k)$, 
with $\mu_k$ solutions of the Bethe equations. Crucially, the $Q$ operator commutes 
with the transfer matrices $T_j(\mu)$, and it can be used to 
construct a solution of the Hirota equations~\eqref{eq:hirota} (see Ref.~\cite{ilievski2016string}). 
In particular, by expanding the solution of~\eqref{eq:hirota} in terms of the $Q$ operator, 
it has been shown that in the thermodynamic limit $L\to\infty$ one has that~\cite{ilievski2016quasilocal}  
\begin{equation}
	\label{eq:T-approx}
	\lim_{L\gg 1} \frac{T_j^{[+]}(\mu)}{T_0^{[j+1]}(\mu)}=\frac{Q^{[-j]}(\mu)}{Q^{[+j]}\mu}=
	\prod_{k=1}^{M_j}\frac{\sin(\mu-\mu_k-j\eta/2)}{\sin(\mu-\mu_k+j\eta/2)}=
	\prod_{k=1}^{M_j} S_j(\mu-\mu_k), 
\end{equation}
where we defined the scattering matrix $S_j(\mu-\mu_k)$ between a magnon and a $j$-particle bound state, 
and $M_j$ is the number of $j$-particle bound states in the Bethe state. 
Importantly, we can now employ Eq.~\eqref{eq:T-approx} to compute 
the expectation value of the local and quasilocal conserved quantities $X_j(\mu)$ (cf.~\eqref{eq:X}).  Specifically, 
from~\eqref{eq:X} and~\eqref{eq:T-approx} we obtain 
\begin{equation}
	X_j(\mu)=\frac{1}{L}\frac{1}{2\pi i}\sum_{k=1}^{M_j}\partial_\mu \ln(S_j(\mu-\mu_k)), 
\end{equation}
where $S_j(\mu-\mu_k)$ is defined in~\eqref{eq:T-approx}. Notice also that $\partial_{\mu}\ln(S_j(\mu-\mu_k))$ is 
defined in~\eqref{eq:dTheta}. 
Following the standard $TBA$ approach~\cite{takahashi1999thermodynamics}, 
in the thermodynamic limit we can replace the sum over the rapidities $\mu_k$ with an integral 
weighted with the root density $\rho_k$ as 
\begin{multline}
	\label{eq:X-rho}
	X_j(\mu)=\int_{-\pi/2}^{\pi/2}\frac{d\lambda}{2\pi i}\sum_{k=1}^{M_j}\sum_{l=1}^k
	\partial_\mu\ln\left[S_j(\mu-\lambda+(k+1-2l)i\eta/2)\right]\rho_k(\lambda):=\\
	\int_{-\pi/2}^{\pi/2}d\lambda \sum_{k=1}^{M_j}G_{j,k}(\mu-\lambda)\rho_k(\lambda)=\sum_{k=1}^{M_j}G_{j,k}\star \rho_k, 
\end{multline}
where the $\star$ symbol denotes convolution (cf.~\eqref{eq:conv}). Crucially, it can be shown that~\cite{ilievski2016string} 
the kernel $G_{j,k}$ is the Green's function for the discrete Laplacian operator $\Box$, which acts on a generic function 
$f_j$ as 
\begin{equation}
	\label{eq:box-def}
	\Box f_j:=f_j^{[+]}+f_j^{[-]}-f_{j+1}-f_{j-1}, \quad \left(\Box G\right)_{j,k}=\delta_{j,k}\delta(\lambda), 
\end{equation}
where $\delta(\lambda)$ is a Dirac delta. Now, we can invert~\eqref{eq:X-rho} by applying the 
Laplacian on both sides, to obtain 
\begin{equation}
	\label{eq:rho-X}
	\rho_n(\mu)=\Box X_n(\mu), 
\end{equation}
where $\Box X_n(\mu)$ is defined in~\eqref{eq:box-def}. 
Eq.~\eqref{eq:rho-X} states that one can obtain the densities $\rho_n$ describing a thermodynamic 
macrostate (such as the $GGE$ arising after a quench) starting from the expectation values of the 
quasilocal charges $X_n$. By using the $TBA$ equation~\eqref{eq:rho-decoup}, 
one can obtain the analog of~\eqref{eq:rho-X} for the hole densities $\rho_{h,n}$ as 
\begin{equation}
	\label{eq:X-rhoh}
	\rho_{h,n}=a_n-X_n^{[+]}-X_n^{[-]}, 
\end{equation}
where we employ the definition in~\eqref{eq:shift}, $X_n$ is the expectation value of the 
generic quasilocal conserved quantity on the initial state, and $a_n$ is the same as in~\eqref{eq:a-def}. 

To proceed, one has to determine the expectation value of both members of~\eqref{eq:X} 
over the initial state $\psi_0$.  A major complication is that Eq.~\eqref{eq:X} requires the inverse 
of the transfer matrix $T_j^{[+]}(\mu)$, which is a daunting task in general. 
Fortunately, in the large $L$ limit one can show that~\cite{ilievski2016quasilocal} 
\begin{equation}
	\label{eq:inversion}
	\lim_{L\gg 1} \frac{T^{[-]}_j(\mu) T_j^{[+]}(\mu)}{{T_0^{[-j-1]}}(\mu)T_0^{[j+1]}(\lambda)}=1.  
\end{equation}
Eq.~\eqref{eq:inversion} provides an inversion formula. This  is crucial to rewrite the logarithmic 
derivative in~\eqref{eq:X} as 
\begin{equation}
	\label{eq:inv-appr}
	X_j(\mu)\simeq \frac{1}{L}\frac{1}{2\pi i}\frac{T_j^{[-]}(\mu)}{T_0^{[-j-1]}(\mu)}\partial_\mu\frac{T_j^{[+]}(\mu)}{T_0^{[j+1]}(\mu)}, 
\end{equation}
which holds in the large $L$ limit. 

We are now ready to compute the expectation value of $X_j$ over the initial 
state $\psi_0$. Here we consider initial states $\psi_0$ that are obtained by repeating a 
unit cell of length $q=2p$, as it is the case for the $p$-N\'eel states~\eqref{eq:pneel}. 
This allows us to rewrite~\eqref{eq:inv-appr} as 
\begin{equation}
	\label{eq:trace-X}
	X_j^{\psi_0}(\mu)=\frac{1}{L}\frac{1}{2\pi i}\mathrm{Tr}_{V_j}\left[
	\partial_x\left.\left(\langle\psi_0^q|
	\prod_{k=1}^{N_q}\frac{L_j^{(k),[-]}(\mu)}{L_0^{(k),[-j-1]}(\mu)}\frac{L_j^{(k),[+]}(\mu+x)}{L_0^{(k),[j+1]}(\mu+x)}|\psi_0^q\rangle
	\right)\right|_{x=0}\right]^{L/L_q}, 
\end{equation}
where we denoted with $X_j^{\psi_0}$ the expectation value of the charges $X_j$ over the initial 
state, and with $|\psi_0^q\rangle$ the unit cell that is used to build $|\psi_0\rangle=|\psi_0^q\rangle^{\otimes L/L_q}$. 
In~\eqref{eq:trace-X} the Lax operators $L_j^{(k),[\pm]}$ are 
obtained from the original ones in~\eqref{eq:lax} by using~\eqref{eq:shift}, and the trace is over the 
auxiliary Hilbert space. Notice that the expectation value over the state $|\psi^q_0\rangle$ of the unit cell 
involves the physical Hilbert space. Finally, in~\eqref{eq:trace-X} we shifted the 
argument of the second term in the brackets to 
bring the derivative outside of the expectation value. 
Eq.~\eqref{eq:trace-X}, together with Eq.~\eqref{eq:rho-X} and Eq.~\eqref{eq:X-rhoh}, is  the main tool 
to  determine the $GGE$. It is useful to observe that Eq.~\eqref{eq:trace-X} can be rewritten as 
\begin{equation}
	\label{eq:trace-X-1}
	X_j^{\psi_0}(\mu)=\frac{1}{L}\frac{1}{2\pi i}\left[\mathrm{Tr}_{V_j}\left.\left(\partial_x
	\mathbb{T}_j^{\psi_0^q}(x,\mu)\right|_{x=0}\right)\right], 
\end{equation}
where we defined $\mathbb{T}^{\psi_0^q}_j$ as 
\begin{equation}
	\mathbb{T}_j^{\psi_0^q}(x,\mu):=\langle\psi_0^q|\mathbb{L}_j^{(N_q)}(x,\mu)\mathbb{L}^{(N_q-1)}\cdots \mathbb{L}^{(1)}_j(x,\mu)|\psi_0^q\rangle, 
\end{equation}
with 
\begin{equation}
	\mathbb{L}_j^{(k)}(x,\mu)=\frac{L_j^{(k),[-]}(\mu) L_j^{(k),[+]}(\mu+x)}{L_0^{(k),[-j-1]}(\mu) L_0^{(k),[j+1]}(\mu+x)}. 
\end{equation}

\section{Convergence of $tDMRG$} 
\label{sec:bond}

\begin{figure}[t]
	\begin{center}
\includegraphics[width=.6\textwidth]{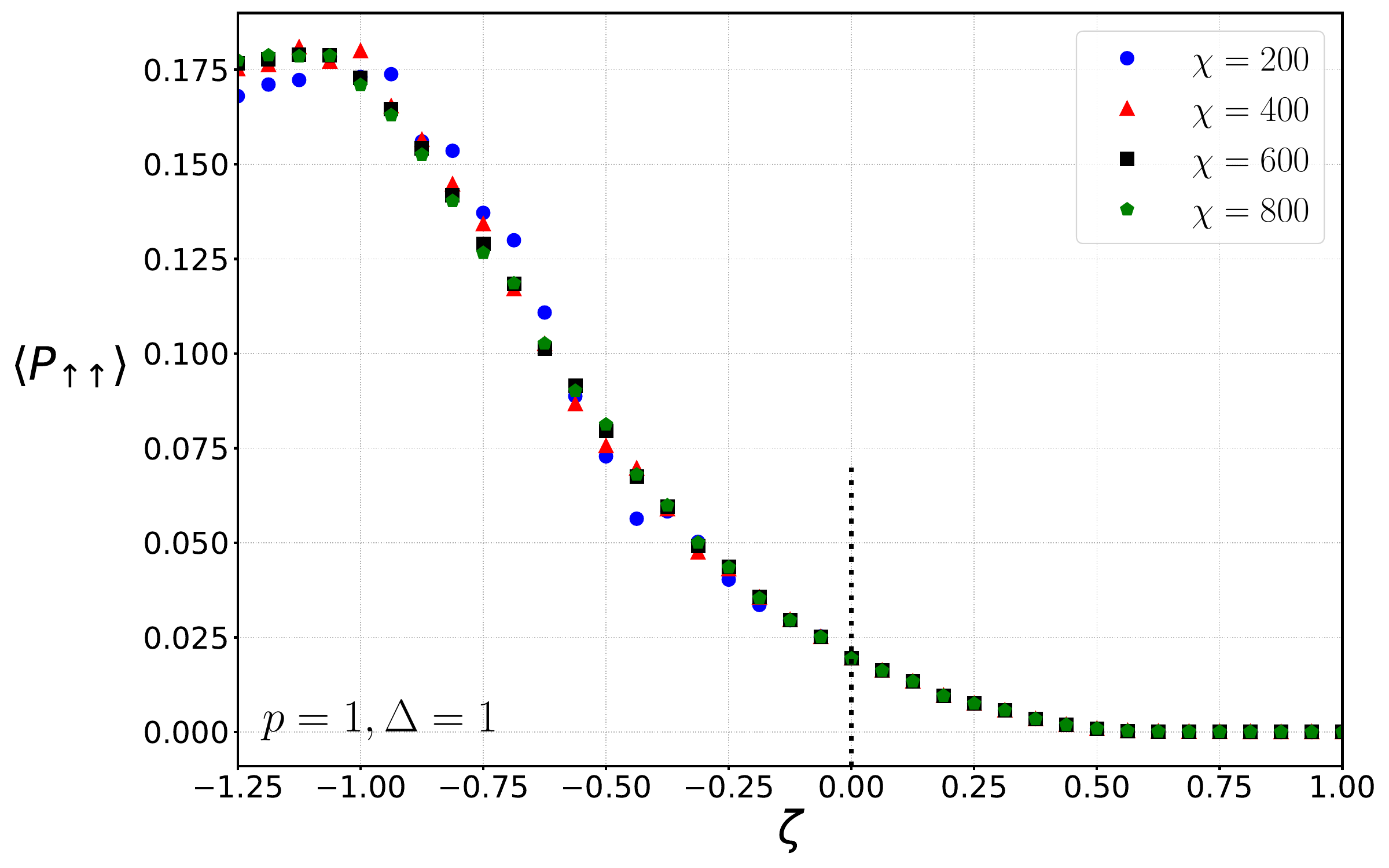}
\caption{ Convergence of the $tDMRG$ for the full-expansion dynamics. We plot the projector $P_{\uparrow\uparrow}$ 
 as a function of $\zeta=x/t$. We focus on the $XXZ$ chain with  $\Delta=1$ and 
 the expansion of the N\'eel state. The initial region $A$ (see Fig.~\ref{fig0:quench}) 
 contains $\ell=40$ sites. The different symbols corresponds to different 
 values of the bond dimension $\chi$. All the results are for $t=16$. 
}
\label{fig:chi}
\end{center}
\end{figure}

Here we discuss the convergence of the $tDMRG$ algorithm. We consider the expansion of the 
N\'eel state, i.e., the $p$-N\'eel state  with $p=1$ focussing on the $XXZ$ chain with $\Delta=1$. 
Typically, we observe that the choice $\Delta=1$, $p=1$ gives rise to the strongest 
entanglement growth. We discuss the convergence of the projector $P_{x,\uparrow\uparrow}$, 
as it is relevant to assess the confinement of the two-particle bound states. In 
Fig.~\ref{fig:chi} we plot the time-dependent expectation value 
$\langle P_{x,\uparrow\uparrow}\rangle$ versus $\zeta=x/t$. The region $A$ prepared 
in the N\'eel state has $\ell=40$ sites. We show data at fixed time $t=16$. In our 
$tDMRG$ simulations we employ a seventh-order Trotter decomposition (see~\cite{bidzhiev-2017}) of the 
evolution operator, with time step $\delta t=0.2$. 
The different symbols in Fig.~\ref{fig:chi} correspond to different values of the bond 
dimension $\chi$ up to $\chi=800$. 
As it is clear from Fig.~\ref{fig:chi} for $\zeta<0$, the finite-$\chi$ effects 
are visible, decreasing with increasing $\chi$. In particular, already 
at $\chi=800$ the convergence is satisfactory, and even at $\chi=200$ the 
qualitative behavior of the data is correctly captured. Finally, for $\zeta>0$ 
finite-$\chi$ effects are significantly smaller. This is expected because entanglement, 
which limits the efficiency of $tDMRG$,  is produced in region $A$, whereas it is 
simply ``transported''~\cite{alba2021generalized} outside of $A$. This is the reason why with $\chi\approx 1000$, 
we can reach $t\approx 100$ in our simulations.

\section*{References}
\bibliographystyle{iopart-num.bst}
\bibliography{bibliography.bib}

\providecommand{\newblock}{}
\begin{thebibliography}{10}
\expandafter\ifx\csname url\endcsname\relax
  \def\url#1{{\tt #1}}\fi
\expandafter\ifx\csname urlprefix\endcsname\relax\def\urlprefix{URL }\fi
\providecommand{\eprint}[2][]{\url{#2}}

\bibitem{bethe1931zur}
Bethe H 1931 {\em Zeitschrift f{\"u}r Physik\/} {\bf 71} 205--226 ISSN
  0044-3328 \urlprefix\url{https://doi.org/10.1007/BF01341708}

\bibitem{takahashi1999thermodynamics}
Takahashi M 1999 {\em Thermodynamics of One-Dimensional Solvable Models\/}
  (Cambridge University Press)

\bibitem{haller2009realization}
Haller E, Gustavsson M, Mark M~J, Danzl J~G, Hart R, Pupillo G and Nägerl H~C
  2009 {\em Science\/} {\bf 325} 1224--1227 (\textit{Preprint}
  \eprint{https://www.science.org/doi/pdf/10.1126/science.1175850})
  \urlprefix\url{https://www.science.org/doi/abs/10.1126/science.1175850}

\bibitem{fukuhara2013microscopic}
Fukuhara T, Schau{\ss} P, Endres M, Hild S, Cheneau M, Bloch I and Gross C 2013
  {\em Nature\/} {\bf 502} 76--79 ISSN 1476-4687
  \urlprefix\url{https://doi.org/10.1038/nature12541}

\bibitem{wang2018experimental}
Wang Z, Wu J, Yang W, Bera A~K, Kamenskyi D, Islam A~T~M~N, Xu S, Law J~M, Lake
  B, Wu C and Loidl A 2018 {\em Nature\/} {\bf 554} 219--223 ISSN 1476-4687
  \urlprefix\url{https://doi.org/10.1038/nature25466}

\bibitem{bera2020dispersion}
Bera A~K, Wu J, Yang W, Bewley R, Boehm M, Xu J, Bartkowiak M, Prokhnenko O,
  Klemke B, Islam A~T~M~N, Law J~M, Wang Z and Lake B 2020 {\em Nature
  Physics\/} {\bf 16} 625--630 ISSN 1745-2481
  \urlprefix\url{https://doi.org/10.1038/s41567-020-0835-7}

\bibitem{scheie2022quantum}
Scheie A, Laurell P, Lake B, Nagler S~E, Stone M~B, Caux J~S and Tennant D~A
  2022 {\em Nature Communications\/} {\bf 13} 5796 ISSN 2041-1723
  \urlprefix\url{https://doi.org/10.1038/s41467-022-33571-8}

\bibitem{kranzl2023observation}
Kranzl F, Birnkammer S, Joshi M~K, Bastianello A, Blatt R, Knap M and Roos C~F
  2023 {\em Phys. Rev. X\/} {\bf 13}(3) 031017
  \urlprefix\url{https://link.aps.org/doi/10.1103/PhysRevX.13.031017}

\bibitem{ilievski2021superuniversality}
Ilievski E, De~Nardis J, Gopalakrishnan S, Vasseur R and Ware B 2021 {\em Phys.
  Rev. X\/} {\bf 11}(3) 031023
  \urlprefix\url{https://link.aps.org/doi/10.1103/PhysRevX.11.031023}

\bibitem{morvan2022formation}
Morvan A, Andersen T~I, Mi X, Neill C, Petukhov A, Kechedzhi K, Abanin D~A,
  Michailidis A, Acharya R, Arute F, Arya K, Asfaw A, Atalaya J, Bardin J~C,
  Basso J, Bengtsson A, Bortoli G, Bourassa A, Bovaird J, Brill L, Broughton M,
  Buckley B~B, Buell D~A, Burger T, Burkett B, Bushnell N, Chen Z, Chiaro B,
  Collins R, Conner P, Courtney W, Crook A~L, Curtin B, Debroy D~M, Del
  Toro~Barba A, Demura S, Dunsworth A, Eppens D, Erickson C, Faoro L, Farhi E,
  Fatemi R, Flores~Burgos L, Forati E, Fowler A~G, Foxen B, Giang W, Gidney C,
  Gilboa D, Giustina M, Grajales~Dau A, Gross J~A, Habegger S, Hamilton M~C,
  Harrigan M~P, Harrington S~D, Hoffmann M, Hong S, Huang T, Huff A, Huggins
  W~J, Isakov S~V, Iveland J, Jeffrey E, Jiang Z, Jones C, Juhas P, Kafri D,
  Khattar T, Khezri M, Kieferov{\'a} M, Kim S, Kitaev A~Y, Klimov P~V, Klots
  A~R, Korotkov A~N, Kostritsa F, Kreikebaum J~M, Landhuis D, Laptev P, Lau
  K~M, Laws L, Lee J, Lee K~W, Lester B~J, Lill A~T, Liu W, Locharla A, Malone
  F, Martin O, McClean J~R, McEwen M, Meurer~Costa B, Miao K~C, Mohseni M,
  Montazeri S, Mount E, Mruczkiewicz W, Naaman O, Neeley M, Nersisyan A, Newman
  M, Nguyen A, Nguyen M, Niu M~Y, O'Brien T~E, Olenewa R, Opremcak A, Potter R,
  Quintana C, Rubin N~C, Saei N, Sank D, Sankaragomathi K, Satzinger K~J,
  Schurkus H~F, Schuster C, Shearn M~J, Shorter A, Shvarts V, Skruzny J, Smith
  W~C, Strain D, Sterling G, Su Y, Szalay M, Torres A, Vidal G, Villalonga B,
  Vollgraff-Heidweiller C, White T, Xing C, Yao Z, Yeh P, Yoo J, Zalcman A,
  Zhang Y, Zhu N, Neven H, Bacon D, Hilton J, Lucero E, Babbush R, Boixo S,
  Megrant A, Kelly J, Chen Y, Smelyanskiy V, Aleiner I, Ioffe L~B and Roushan P
  2022 {\em Nature\/} {\bf 612} 240--245 ISSN 1476-4687
  \urlprefix\url{https://doi.org/10.1038/s41586-022-05348-y}

\bibitem{surace2024robustness}
Surace F~M and Motrunich O 2024 {\em PRX Quantum\/} {\bf 5}(1) 010317
  \urlprefix\url{https://link.aps.org/doi/10.1103/PRXQuantum.5.010317}

\bibitem{alba2017entanglement}
Alba V and Calabrese P 2017 {\em Proceedings of the National Academy of
  Sciences\/} {\bf 114} 7947--7951 ISSN 0027-8424 (\textit{Preprint}
  \eprint{https://www.pnas.org/content/114/30/7947.full.pdf})
  \urlprefix\url{https://www.pnas.org/content/114/30/7947}

\bibitem{bertini2016transport}
Bertini B, Collura M, De~Nardis J and Fagotti M 2016 {\em Phys. Rev. Lett.\/}
  {\bf 117}(20) 207201
  \urlprefix\url{https://link.aps.org/doi/10.1103/PhysRevLett.117.207201}

\bibitem{castro-alvaredo2016emergent}
Castro-Alvaredo O~A, Doyon B and Yoshimura T 2016 {\em Phys. Rev. X\/} {\bf
  6}(4) 041065
  \urlprefix\url{https://link.aps.org/doi/10.1103/PhysRevX.6.041065}

\bibitem{calabrese2016introduction}
Calabrese P, Essler F~H~L and Mussardo G 2016 {\em Journal of Statistical
  Mechanics: Theory and Experiment\/} {\bf 2016} 064001
  \urlprefix\url{https://doi.org/10.1088/1742-5468/2016/06/064001}

\bibitem{alba2018entanglementand}
Alba V 2018 {\em Phys. Rev. B\/} {\bf 97}(24) 245135
  \urlprefix\url{https://link.aps.org/doi/10.1103/PhysRevB.97.245135}

\bibitem{alba2019towards}
Alba V 2019 {\em Phys. Rev. B\/} {\bf 99}(4) 045150
  \urlprefix\url{https://link.aps.org/doi/10.1103/PhysRevB.99.045150}

\bibitem{alba2019entanglement}
Alba V, Bertini B and Fagotti M 2019 {\em SciPost Phys.\/} {\bf 7}(1) 5
  \urlprefix\url{https://scipost.org/10.21468/SciPostPhys.7.1.005}

\bibitem{alba2021generalized}
Alba V, Bertini B, Fagotti M, Piroli L and Ruggiero P 2021 {\em Journal of
  Statistical Mechanics: Theory and Experiment\/} {\bf 2021} 114004
  \urlprefix\url{https://doi.org/10.1088/1742-5468/ac257d}

\bibitem{piroli2017transport}
Piroli L, De~Nardis J, Collura M, Bertini B and Fagotti M 2017 {\em Phys. Rev.
  B\/} {\bf 96}(11) 115124
  \urlprefix\url{https://link.aps.org/doi/10.1103/PhysRevB.96.115124}

\bibitem{heidrich-meisner2009quantum}
Heidrich-Meisner F, Manmana S~R, Rigol M, Muramatsu A, Feiguin A~E and Dagotto
  E 2009 {\em Phys. Rev. A\/} {\bf 80}(4) 041603
  \urlprefix\url{https://link.aps.org/doi/10.1103/PhysRevA.80.041603}

\bibitem{xia2015quantum}
Xia L, Zundel L~A, Carrasquilla J, Reinhard A, Wilson J~M, Rigol M and Weiss
  D~S 2015 {\em Nature Physics\/} {\bf 11} 316--320 ISSN 1745-2481
  \urlprefix\url{https://doi.org/10.1038/nphys3244}

\bibitem{scherg2018nonequilibrium}
Scherg S, Kohlert T, Herbrych J, Stolpp J, Bordia P, Schneider U,
  Heidrich-Meisner F, Bloch I and Aidelsburger M 2018 {\em Phys. Rev. Lett.\/}
  {\bf 121}(13) 130402
  \urlprefix\url{https://link.aps.org/doi/10.1103/PhysRevLett.121.130402}

\bibitem{paeckel2019time}
Paeckel S, Köhler T, Swoboda A, Manmana S~R, Schollw\"ock U and Hubig C 2019
  {\em Annals of Physics\/} {\bf 411} 167998 ISSN 0003-4916
  \urlprefix\url{https://www.sciencedirect.com/science/article/pii/S0003491619302532}

\bibitem{ilievski2016string}
Ilievski E, Quinn E, Nardis J~D and Brockmann M 2016 {\em Journal of
  Statistical Mechanics: Theory and Experiment\/} {\bf 2016} 063101
  \urlprefix\url{https://dx.doi.org/10.1088/1742-5468/2016/06/063101}

\bibitem{ilievski2016quasilocal}
Ilievski E, Medenjak M, Prosen T and Zadnik L 2016 {\em Journal of Statistical
  Mechanics: Theory and Experiment\/} {\bf 2016} 064008
  \urlprefix\url{https://dx.doi.org/10.1088/1742-5468/2016/06/064008}

\bibitem{ganahl2012observation}
Ganahl M, Rabel E, Essler F~H~L and Evertz H~G 2012 {\em Phys. Rev. Lett.\/}
  {\bf 108}(7) 077206
  \urlprefix\url{https://link.aps.org/doi/10.1103/PhysRevLett.108.077206}

\bibitem{calabrese2005evolution}
Calabrese P and Cardy J 2005 {\em Journal of Statistical Mechanics: Theory and
  Experiment\/} {\bf 2005} P04010
  \urlprefix\url{https://doi.org/10.1088{\%}2F1742-5468{\%}2F2005{\%}2F04{\%}2Fp04010}

\bibitem{fagotti2008evolution}
Fagotti M and Calabrese P 2008 {\em Phys. Rev. A\/} {\bf 78}(1) 010306
  \urlprefix\url{https://link.aps.org/doi/10.1103/PhysRevA.78.010306}

\bibitem{gobert-2005}
Gobert D, Kollath C, Schollw\"ock U and Sch\"utz G 2005 {\em Phys. Rev. E\/}
  {\bf 71}(3) 036102
  \urlprefix\url{https://link.aps.org/doi/10.1103/PhysRevE.71.036102}

\bibitem{ilievski2018superdiffusion}
Ilievski E, De~Nardis J, Medenjak M and Prosen T 2018 {\em Phys. Rev. Lett.\/}
  {\bf 121}(23) 230602
  \urlprefix\url{https://link.aps.org/doi/10.1103/PhysRevLett.121.230602}

\bibitem{bonnes2014light}
Bonnes L, Essler F~H~L and L\"auchli A~M 2014 {\em Phys. Rev. Lett.\/} {\bf
  113}(18) 187203
  \urlprefix\url{https://link.aps.org/doi/10.1103/PhysRevLett.113.187203}

\bibitem{korepin1993quantum}
Korepin V~E, Bogoliubov N~M and Izergin A~G 1993 {\em Quantum Inverse
  Scattering Method and Correlation Functions\/} Cambridge Monographs on
  Mathematical Physics (Cambridge University Press)

\bibitem{bulchandani2018bethe}
Bulchandani V~B, Vasseur R, Karrasch C and Moore J~E 2018 {\em Phys. Rev. B\/}
  {\bf 97}(4) 045407
  \urlprefix\url{https://link.aps.org/doi/10.1103/PhysRevB.97.045407}

\bibitem{calabrese-2005}
Calabrese P and Cardy J 2005 {\em Journal of Statistical Mechanics: Theory and
  Experiment\/} {\bf 2005} P04010
  \urlprefix\url{https://doi.org/10.1088{\%}2F1742-5468{\%}2F2005{\%}2F04{\%}2Fp04010}

\bibitem{alba2018entanglement}
Alba V and Calabrese P 2018 {\em SciPost Phys.\/} {\bf 4}(3) 17
  \urlprefix\url{https://scipost.org/10.21468/SciPostPhys.4.3.017}

\bibitem{schemmer2019generalized}
Schemmer M, Bouchoule I, Doyon B and Dubail J 2019 {\em Phys. Rev. Lett.\/}
  {\bf 122}(9) 090601
  \urlprefix\url{https://link.aps.org/doi/10.1103/PhysRevLett.122.090601}

\bibitem{bouchoule2022generalized}
Bouchoule I and Dubail J 2022 {\em Journal of Statistical Mechanics: Theory and
  Experiment\/} {\bf 2022} 014003
  \urlprefix\url{https://dx.doi.org/10.1088/1742-5468/ac3659}

\bibitem{WeiDavid2022Qgmo}
Wei D, Rubio-Abadal A, Ye B, Machado F, Kemp J, Srakaew K, Hollerith S, Rui J,
  Gopalakrishnan S, Yao N~Y, Bloch I and Zeiher J 2022 {\em Science (American
  Association for the Advancement of Science)\/} {\bf 376} 716--720 ISSN
  0036-8075

\bibitem{macri2021bound}
Macr\`{\i} T, Lepori L, Pagano G, Lewenstein M and Barbiero L 2021 {\em Phys.
  Rev. B\/} {\bf 104}(21) 214309
  \urlprefix\url{https://link.aps.org/doi/10.1103/PhysRevB.104.214309}

\bibitem{vanicat2018integrable}
Vanicat M, Zadnik L and Prosen T 2018 {\em Phys. Rev. Lett.\/} {\bf 121}(3)
  030606
  \urlprefix\url{https://link.aps.org/doi/10.1103/PhysRevLett.121.030606}

\bibitem{ljubotina2019ballistic}
Ljubotina M, Zadnik L and Prosen T 2019 {\em Phys. Rev. Lett.\/} {\bf 122}(15)
  150605
  \urlprefix\url{https://link.aps.org/doi/10.1103/PhysRevLett.122.150605}

\bibitem{claeys2022correlations}
Claeys P~W, Herzog-Arbeitman J and Lamacraft A 2022 {\em SciPost Phys.\/} {\bf
  12} 007 \urlprefix\url{https://scipost.org/10.21468/SciPostPhys.12.1.007}

\bibitem{bazhanov2010a}
Bazhanov V~V, Łukowski T, Meneghelli C and Staudacher M 2010 {\em Journal of
  Statistical Mechanics: Theory and Experiment\/} {\bf 2010} P11002
  \urlprefix\url{https://dx.doi.org/10.1088/1742-5468/2010/11/P11002}

\bibitem{bidzhiev-2017}
Bidzhiev K and Misguich G 2017 {\em Phys. Rev. B\/} {\bf 96}(19) 195117
  \urlprefix\url{https://link.aps.org/doi/10.1103/PhysRevB.96.195117}

\end{thebibliography}

\end{document}